\documentclass[aps,prd,10pt,nofootinbib,showpacs,showkeys,superscriptaddress,preprintnumbers,altaffilletter,floatfix]{revtex4-2}

\pdfoutput=1
\usepackage{soul}
\usepackage{graphicx}
\usepackage{microtype}
\usepackage{amsmath}
\usepackage{amssymb}
\usepackage{subfigure}
\usepackage{hyperref}
\usepackage{url}
\usepackage{xcolor}
\usepackage{color}
\usepackage{mathrsfs}
\usepackage{calrsfs}
\usepackage{amsfonts}
\usepackage{latexsym}
\usepackage{ragged2e}
\usepackage{epsfig}
\usepackage{textcomp}
\usepackage{phaistos}
\usepackage[utf8]{inputenc}
\makeatletter
\renewcommand\@makefnmark{\hbox{\@textsuperscript{\normalfont\color{purple}\@thefnmark}}}
\renewcommand\@makefntext[1]{%
  \parindent 1em\noindent
            \hb@xt@1.8em{%
                \hss\@textsuperscript{\normalfont\@thefnmark}}#1}
\makeatother

\usepackage{verbatim}
\usepackage{caption}
\DeclareCaptionJustification{justified}{\leftskip=0pt \rightskip=0pt \parfillskip=0pt plus 1fil}
\definecolor{vividviolet}{rgb}{0.62, 0.0, 1.0}
\definecolor{amaranth}{rgb}{0.9, 0.17, 0.31}
\definecolor{palatinateblue}{rgb}{0.15, 0.23, 0.89}
\definecolor{brightpink}{rgb}{1.0, 0.0, 0.5}
\definecolor{cornflowerblue}{rgb}{0.39, 0.58, 0.93}
\definecolor{deepcarminepink}{rgb}{0.94, 0.19, 0.22}
\definecolor{radicalred}{rgb}{1.0, 0.21, 0.37}
\definecolor{darkgreen}{rgb}{0.06 0.64, 0.43}
\definecolor{darkpink}{RGB}{231,84,128}

\hypersetup{ linktoc=all,
    colorlinks, linkcolor={palatinateblue},
    citecolor={brightpink}, urlcolor={amaranth}
}

\graphicspath{{Images/}}

\def\sideremark#1{\ifvmode\leavevmode\fi\vadjust{\vbox to0pt{\vss
 \hbox to 0pt{\hskip\hsize\hskip1em
 \vbox{\hsize1.5cm\tiny\raggedright\pretolerance10000
 \noindent #1\hfill}\hss}\vbox to8pt{\vfil}\vss}}}

\usepackage{graphicx} 
\usepackage{comment}

\usepackage{cancel}

\newcommand{\be}{\begin{equation}}
\newcommand{\ee}{\end{equation}}
\newcommand{\bea}{\begin{eqnarray}}
\newcommand{\eea}{\end{eqnarray}}

\newcommand{\bseq}{\begin{subequations}}
\newcommand{\eseq}{\end{subequations}}

\newcommand{\dd}{\mathrm{d}}

\usepackage[]{xcolor}
\usepackage{hyperref}

\usepackage{amsmath}

\begin{document}

\title{Insights in $f(Q)$ cosmology: the relevance of the connection}

\author{Ismael \surname{Ayuso}}
\email{ismael.ayuso@ehu.eus}
\affiliation{Department of Physics \& EHU Quantum Center, University of the Basque Country UPV/EHU, Bilbao 48080, Spain}

\author{Mariam \surname{Bouhmadi-L\'opez}}
\email{mariam.bouhmadi@ehu.eus}
\affiliation{IKERBASQUE, Basque Foundation for Science, Bilbao 48011, Spain}
\affiliation{Department of Physics \& EHU Quantum Center, University of the Basque Country UPV/EHU, Bilbao 48080, Spain}

\author{Che-Yu \surname{Chen}}
\email{b97202056@gmail.com}
\affiliation{RIKEN iTHEMS, Wako, Saitama 351-0198, Japan}

\author{\\ Xiao Yan \surname{Chew}}
\email{xiao.yan.chew@just.edu.cn}
\affiliation{School of Science, Jiangsu University of Science and Technology, 212100, Zhenjiang, China}

\author{Konstantinos \surname{Dialektopoulos}}
\email{kdialekt@gmail.com}
\affiliation{Department of Mathematics and Computer Science, Transilvania University of Brasov, 500091, Brasov, Romania}
\affiliation{Institute of Space Sciences and Astronomy, University of Malta, Msida, Malta and Department of Physics, University of Malta, Msida, Malta}

\author{Yen Chin \surname{Ong}}
\email{ycong@yzu.edu.cn}
\affiliation{Center for Gravitation and Cosmology, College of Physical Science and Technology, Yangzhou University, \\180 Siwangting Road, Yangzhou City, Jiangsu Province  225002, China}
\affiliation{School of Aeronautics and Astronautics, Shanghai Jiao Tong University, Shanghai 200240, China}

\begin{abstract} 
We explore the role of the affine connection in $f(Q)$ gravity, a modified theory where gravity is governed by non-metricity within the symmetric teleparallel framework. Although the connection is constrained to be flat and torsionless, it is not uniquely determined by the metric, allowing for multiple physically distinct formulations. We analyze three such connections compatible with a homogeneous and isotropic universe to show that they yield markedly different cosmological dynamics, even under the same functional form of $f(Q)$. Using both analytical and numerical methods, including a Born-Infeld type model of $f(Q)$, we demonstrate that specific connections can resolve cosmological singularities like the Big Bang and Big Rip, replacing them with smooth de Sitter phases. Others retain singularities but with notable modifications in their behavior. These findings highlight the physical relevance of connection choice in $f(Q)$ gravity and its potential to address fundamental cosmological questions.
\end{abstract}

\maketitle

\section{Introduction}

The standard model of cosmology, known as the $\Lambda$CDM model, has been remarkably successful in explaining a wide array of cosmological observations. It provides an excellent fit to the cosmic microwave background (CMB) anisotropies \cite{Planck:2018vyg}, the large-scale distribution of galaxies, the luminosity distances of Type Ia supernovae \cite{SupernovaSearchTeam:1998fmf}, and the imprint of baryon acoustic oscillations (BAO) across different redshifts. Furthermore, $\Lambda$CDM underpins the current understanding of structure formation and has enabled precision cosmology through large-scale surveys such as SDSS and DES \cite{SDSS:2014iwm,DES:2021wwk}. However, despite these achievements, the model remains conceptually incomplete and faces a number of persistent theoretical and observational challenges. It posits the existence of cold dark matter and a cosmological constant $\Lambda$—components that dominate the universe’s energy content yet lack a well-established physical explanation \cite{Bertone:2016nfn}. The cosmological constant problem, which reveals a profound discrepancy between the observed value of $\Lambda$ and theoretical expectations from quantum field theory, remains unresolved \cite{Weinberg:1988cp,Martin:2012bt}. Moreover, recent high-precision observations have brought certain tensions into sharper focus. These include the Hubble tension, a statistically significant discrepancy between local measurements of the Hubble constant and values inferred from the early universe \cite{Riess:2020fzl,Verde:2019ivm,DiValentino:2025sru}, and the $S_8$ tension, related to the amplitude of matter clustering \cite{Hildebrandt:2018yau}. Notably, the latest results from the Dark Energy Spectroscopic Instrument (DESI) have added further pressure on the standard model by reporting deviations in the growth rate of structure and the expansion history that may be difficult to reconcile within the $\Lambda$CDM framework alone \cite{DESI:2025zgx}. These tensions have motivated the exploration of alternative theories of gravity and dark energy that may offer deeper insight into the fundamental workings of the cosmos.

These shortcomings have led to a growing interest in modified theories of gravity as alternatives or extensions to General Relativity (GR) \cite{Clifton:2011jh}. The motivation behind such modifications is twofold: to explain the observed acceleration of the universe without invoking dark energy, and to explore new theoretical avenues that could bridge the gap between GR and quantum gravity \cite{Joyce:2014kja}. Among the landscape of modified theories, a particularly intriguing class involves changing the geometric foundations of gravity itself. Instead of relying solely on Riemannian geometry and curvature as in GR, these theories utilize alternative geometric quantities such as torsion and non-metricity \cite{Aldrovandi:2013wha,Heisenberg:2018vsk}. This shift in perspective forms the basis of the so-called ``geometrical trinity of gravity''\footnote{There are some concerns mentioned in \cite{Golovnev:2024lku}, where the author argues that the geometric trinity of gravity perspective misrepresents the foundational role of the Levi-Civita connection in GR, asserting that introducing alternative geometric structures without altering physical predictions adds unnecessary complexity. He emphasizes that while these alternative formulations may offer new mathematical frameworks, they do not provide additional physical insights beyond those already encompassed by GR.}, which includes formulations based on curvature (as in GR), torsion (teleparallel gravity), and non-metricity (symmetric teleparallel gravity) \cite{BeltranJimenez:2019esp}. See also Refs.~\cite{Capozziello:2022zzh,Capozziello:2023vne}.

Symmetric teleparallel gravity recasts gravitational dynamics in terms of the non-metricity tensor rather than curvature or torsion \cite{Nester:1998mp}. In this framework, the Levi-Civita connection is replaced by a symmetric and flat connection, and gravity is attributed to the non-metricity scalar $Q$ \cite{BeltranJimenez:2017vop}. This approach preserves the equivalence with GR at the level of field equations when the scalar of curvature of the Einstein-Hilbert action is substituted by $Q$ (the Symmetric Teleparallel Equivalent of General Relativity (STEGR)), but allows for new modifications that can lead to novel phenomenology in other cases. The most studied extension in this context is $f(Q)$ gravity, where the non-metricity scalar $Q$ in the action is generalized to an arbitrary function $f(Q)$ \cite{BeltranJimenez:2018vdo}. Much like $f(R)$ gravity extends GR by promoting the Ricci scalar $R$ to a functional form, $f(Q)$ gravity enables a rich structure of modified dynamics without introducing higher-order derivatives in the field equations, thus avoiding Ostrogradsky instabilities. 

The appeal of $f(Q)$ gravity lies in its mathematical simplicity, second-order field equations, and the geometric reinterpretation of gravity. Over the last few years, $f(Q)$ gravity has been investigated across a variety of domains, including early- and late-time cosmology \cite{Frusciante:2021sio}, inflation \cite{Capozziello:2022tvv, Capozziello:2024lsz} , black hole solutions \cite{Dimakis:2024fan}, relativistic stars \cite{Dimakis:2025jrl}, wormholes \cite{Banerjee:2021mqk}, gravitational wave polarizations \cite{Dong:2025pyz}, and even quantum cosmology \cite{Paliathanasis:2024hvl}. In the cosmological context, $f(Q)$ models can explain cosmic acceleration \cite{Lazkoz:2019sjl, Boiza:2025xpn} and alleviate certain observational tensions \cite{Basilakos:2025olm, Boiza:2025xpn}. In addition, $f(Q)$ allows us to reproduce well-known cosmological scenarios from GR \cite{Chakraborty:2025qlv} and, from there, to build extensions with new phenomenology and compare them with observations \cite{Lazkoz:2019sjl}. Another attempt in this direction is the STEGR supplemented with a general power-law term \cite{BeltranJimenez:2019tme}, which aimed at accounting for the dark energy. However,  such models without cosmological constant are disfavored when confronted with observations \cite{Ayuso:2020dcu, Sahlu:2024pxk}. Another example of this is the DGP model with its extensions being constrained by observational data \cite{Ayuso:2021vtj}. On the other hand, power-law and exponential forms of $f(Q)$ have been shown to yield accelerated expansion consistent with observations \cite{Dialektopoulos:2025ihc,Paliathanasis:2025hjw}.

Despite the progress, several challenges and ambiguities persist in the formulation and application of $f(Q)$ theories. One of the central and underexplored issues is the role of the affine connection. In symmetric teleparallel gravity, the affine connection is not uniquely determined by the metric; instead, there exists a family of connections that satisfy the symmetric and flatness conditions \cite{DAmbrosio:2021pnd,Heisenberg:2023lru}. For instance, in cosmology, while many studies fix the connection to a trivial one for mathematical convenience \cite{BeltranJimenez:2019tme}, recent research has emphasized that different choices of connection can yield distinct physical predictions \cite{Paliathanasis:2023pqp,Shabani:2023xfn,Jarv:2023sbp,Guzman:2024cwa}. This is particularly relevant in $f(Q)$ gravity, where the connection can enter dynamically in the field equations, affecting cosmological evolution and degrees of freedom \cite{Paliathanasis:2024xpy}.

Several studies have now highlighted the dynamical impact of non-trivial connections in $f(Q)$ cosmology. For instance, in \cite{DAmbrosio:2021pnd} the authors demonstrated that some connection components can become genuinely dynamical, leading to new cosmological solutions beyond GR. The review \cite{Heisenberg:2023lru} clarifies further the mathematical underpinnings and implications of the geometrical trinity, emphasizing the promising framework of non-metricity-based theories. Recent works have investigated the appearance of phantom behavior \cite{Shabani:2023xfn}, strong coupling and ghost instabilities  \cite{Gomes:2023tur} that can be solved with non-pathological symmetric teleparallel setup \cite{Bello-Morales:2024vqk}, or a direct coupling of the matter field to the connection or non-minimal couplings \cite{Heisenberg:2023wgk}, tilt-induced anisotropies \cite{Paliathanasis:2023nkb}, and even quantum cosmological corrections via the Wheeler-DeWitt equation \cite{Paliathanasis:2024hvl}, all dependent on the chosen connection. Following the above, in \cite{Gomes:2023tur} and \cite{Saha:2025cfs}, the authors critically assess the viability of $f(Q)$ gravity. Both studies highlight severe theoretical issues, including the presence of ghost degrees of freedom, strong coupling in cosmological perturbations, and the breakdown of standard constraint analysis methods. These pathologies undermine the consistency and predictive power of the theory, especially in early-universe applications. To overcome these limitations, the latter paper proposes extending the framework to $f(R, Q)$ gravity, incorporating both curvature and non-metricity scalars, while \cite{Hu:2023gui} identifies a ghost in scalar-nonmetricity theories but shows it to be nonpropagating via ADM analysis and second-class constraints. However, in \cite{Gomes:2023tur} the authors claim that $f(Q)$ theory propagates the maximum number of degrees of freedom, and this implies there will always be a ghost—not just in cosmological solutions, but in all possible configurations. The remedy comes with \cite{Bello-Morales:2024vqk}, where a class of ghost-free models can be constructed in the framework of symmetric teleparallel geometry with the Lagrangian  $\mathcal{L} = Q + c\, Q_\alpha Q^\alpha$.  This case is special because the term $Q_\alpha Q^\alpha$ at least preserves Transverse Diffs (a subgroup of Diffs corresponding to diffeomorphisms with unit Jacobian). Consequently, it is possible to remove ghosts from the spectrum, as in \cite{Alvarez:2006uu}, where this class of theories has been analysed.

In the black hole and stellar context, $f(Q)$ gravity has been applied to derive new exact solutions, examine horizon structures, and understand matter couplings \cite{Dimakis:2024fan}. The use of Noether symmetries in the minisuperspace approach has allowed the derivation of analytic cosmological and static solutions, particularly in power-law and exponential models \cite{Dialektopoulos:2025ihc,Paliathanasis:2023pqp}. Nevertheless, some of these models exhibit pathologies, such as ghost degrees of freedom and strong coupling, especially in perturbative regimes \cite{Hohmann:2021fpr}, indicating the need for a more careful treatment of the connection.

This growing body of work suggests that the affine connection in $f(Q)$ gravity is not a mere auxiliary construct but a physically meaningful component that directly influences observable predictions. It becomes essential, therefore, to systematically study the cosmological consequences of different symmetric and flat connections, especially in homogeneous and isotropic spacetimes. Understanding the role of the connection may help address some of the unresolved issues in $\Lambda$CDM cosmology and provide a framework for constructing more predictive and robust theories of gravity \cite{Heisenberg:2023lru}.

In this paper, we aim to further investigate the cosmological dynamics of $f(Q)$ gravity by explicitly analyzing the impact of different affine connections compatible with spatial homogeneity and isotropy. By studying various connections within the symmetric teleparallel framework, we demonstrate that the resulting cosmological evolution can vary significantly depending on the choice of connection, even within the same $f(Q)$ model. We explore analytical solutions, including de Sitter spacetimes, and numerically simulate cosmological evolution under a Born-Infeld type $f(Q)$ function. Inspired by its huge success in avoiding field divergences in classical electrodynamics \cite{Born:1934gh}, the Born-Infeld structure has been applied to gravitational sectors in order to address the singularity problems in GR \cite{BeltranJimenez:2017doy}. In the standard Riemannian setup, such structures in the action generically lead to ghosts \cite{Deser:1998rj}. In this regard, the extension to the metric-affine formalism arises, which includes the Palatini approach \cite{Banados:2010ix,Chen:2015eha,BeltranJimenez:2021oaq} and Born-Infeld $f(T)$ gravity in torsional teleparallelism \cite{Ferraro:2006jd,Ferraro:2008ey,Fiorini:2013kba,Bouhmadi-Lopez:2014tna}. In the context of Born-Infeld $f(Q)$ gravity, our findings show that certain connections can regularize the Big Bang and Big Rip singularities, replacing them with smooth de Sitter phases, while others lead to different behaviors such as stiff-matter regimes or phantom-divide crossings. These results underscore the need to treat the affine connection as a central player in the dynamics of modified gravity, opening new directions for resolving foundational problems in cosmology through geometric reformulations of gravity.

The paper is organized as follows: in Sec. \ref{Revf(Q)} we present the symmetric teleparallel geometry and specifically the $f(Q)$ theory. In Sec. \ref{cosmof(Q)} we discuss its cosmology and the three different connections that it can accept and we present the Friedmann equations for each one of them. We show that for a homogeneous and isotropic universe, the equations of motion of the connections have exact integrals that constrain the cosmological evolution. In Sec. \ref{subsec:maximsym} we study analytical cosmological solutions. In particular, we find for which $f(Q)$ models the theory accepts maximally symmetric solutions both in vacuum and in the presence of matter, while we also assume power-law and exponential models and present their solutions for the different connections. In the last section \ref{sec:fqcosmology}, we study numerically the cosmology of Born-Infeld $f(Q)$ theory and we discuss the differences between the connections. We finally conclude in Sec.~\ref{sec:conclustion}.

\section{Reviewing $f(Q)$ gravitational theory} \label{Revf(Q)}

A manifold describing a spacetime can be endowed with a metric $g_{\mu\nu}$ and a connection\footnote{$\Gamma^\alpha_{\;\;\mu\nu}$ is more appropriately called a connection coefficient, but for convenience we refer to connection coefficients as just the ``connection''.} $\Gamma^\alpha_{\;\;\mu\nu}$. On the one  hand, the metric is responsible for measuring the distance between two points. On the other hand, the connection describes the parallel transportation of a given vector. If two vectors can be transported in a commutative way reaching the same point, the connection is symmetric ($\Gamma^\alpha_{\;\;\mu\nu}=\Gamma^\alpha_{\;\;\nu\mu}$). Otherwise, the connection is asymmetric and torsion arises as $T_{\mu\nu}=\Gamma^\alpha_{\;\;[\mu\nu]}$. If the length of a given vector is conserved when parallely transported, then the connection is metric compatible, i.e. $(\nabla_\alpha g_{\mu\nu}=0)$. If the connection is not metric compatible, we  can define the non-metricity tensor as $Q_{\alpha\mu\nu}:=\nabla_\alpha g_{\mu\nu}$. Any connection can be expressed as the sum of the Levi-Civita part (a torsionless and metric compatible connection) together with a term built from torsion and a term arising from non-metricity (cf. Ref.~\cite{Jarv:2018bgs})

For an arbitrary connection, the Riemann tensor is defined as:
\bea
R^{\alpha}_{\;\;\beta\mu\nu}(\Gamma)=2\partial_{[\mu}\Gamma^{\alpha}_{\;\;\nu]\beta}+2\Gamma^{\alpha}_{\;\;[\mu|\lambda|}\Gamma^{\lambda}_{\;\;\nu]\beta}\; .\label{riemann}
\eea
In this way, it is possible to consider a connection $\Gamma^\alpha_{\;\;\mu\nu}$ such that $R^{\alpha}_{\;\;\beta\mu\nu}(\Gamma)=0$. Then, the connection is not \textit{curved} and the spacetime is flat \cite{BeltranJimenez:2019esp}, denoting this case as Teleparallel Gravity. 

Let us consider the Einstein-Hilbert action with an arbitrary connection. We derive the equations of motion  by varying the Einstein-Hilbert action with respect to the metric  and the connection. This approach is known as the metric-affine formalism, which reduces to the Palatini formalism when the matter Lagrangian is independent of the connection. However, this is not GR since the connection does not necessarily have to be the Levi-Civita connection, i.e., torsionless and metric compatible. If we impose the torsionless condition (or metric-compatibility condition), the equations of motion obtained from variation of the Einstein-Hilbert action with respect to the connection will enforce metric compatibility condition (or torsionless condition), and consequently, GR emerges as:
\bea
S=\int{d^4 x\sqrt{-g} R(\{\})}\; ,\label{EHaction}
\eea
where $R(\{\})$ denotes the scalar curvature for the Levi-Civita connection \cite{Clifton:2011jh}. 

Nevertheless, it is possible to rewrite action \eqref{EHaction} using the non-metricity and in the framework of Symmetric Teleparallel Gravity (STG), i.e., for a torsionless and flat connection.  Hence, the following non-metricity scalar is proposed \cite{DAmbrosio:2021pnd}:
\bea
Q:=-\frac{1}{4}Q_{\alpha\mu\nu}Q^{\alpha\mu\nu}+\frac{1}{2}Q_{\alpha\mu\nu}Q^{\mu\alpha\nu}+\frac{1}{4}Q_\alpha Q^\alpha-\frac{1}{2}Q_\alpha \tilde{Q}^\alpha\; ,\nonumber
\eea
where:
\bea
Q_\alpha:=Q_{\alpha\mu}^{\;\;\;\;\mu} \;\;\;\;\; \;\;\;\;\;\tilde{Q}_\alpha:=Q_{\mu\alpha}^{\;\;\;\;\mu}\; .
\eea
Then, from Eq. \eqref{riemann} and the decomposition of a general connection, the following relation appears in STG \cite{Ayuso:2020dcu}:
\bea
R(\{\})=Q-\nabla_{\mu}^{\{\}}\left(Q^\mu-\tilde{Q}^\mu\right)\label{R_EH}\; ,
\eea
where $\nabla_{\mu}^{\{\}}$ is a total divergence term (calculated from the Levi-Civita connection \cite{Ayuso:2020dcu}). Therefore, the second term on the right-hand side (rhs) of Eq. \eqref{R_EH} will be a surface term in action \eqref{EHaction}, playing no role when deriving the equations of motion. As a result, we can describe GR in terms of the non-metricity scalar. However, the equality breaks when we promote this approach to a general function. In fact, $f(R)$ gravity and $f(Q)$ gravity are generally inequivalent for a given function $f$. The same happens for Teleparallel Equivalent of General Relativity (TEGR), where we can rewrite GR using torsion $(T)$, but $f(R)\neq f(T)$. 

Therefore, this presents an area for exploration based on the non-metricity scalar, which could be phenomenologically different from theories constructed from $R(\{\})$. The $f(Q)$ gravitational theories are described by the gravitational action \cite{Khyllep:2021pcu}:
\begin{equation}
    S=\int{d^4 x\sqrt{-g}\left[f(Q)+\mathcal{L}_{M}\right]}\,.
\end{equation}
From now onwards, the units  are adopted as $8\pi G=1$ and $c=1$. We derive the modified Einstein equation by varying the above action with respect to the metric \cite{Xu:2019sbp}: 

\bea
\frac{2}{\sqrt{-g}}\nabla_\alpha\left(\sqrt{-g}f'(Q)P^{\alpha}_{\;\;\;\mu\nu}\right)+\frac{1}{2}g_{\mu\nu}f(Q)+f'(Q)P_{\mu\alpha\beta}Q_{\nu}^{\;\;\;\alpha\beta}=T_{\mu\nu}\; , \label{meteq}
\eea
where $f'\equiv df(Q)/dQ$, and $P^\alpha_{\;\;\;\mu\nu}$ are the components of the non-metricity conjugate tensor defined as:
\bea
P^\alpha_{\;\;\;\mu\nu}=-\frac{1}{2}L^{\alpha}_{\;\;\;\mu\nu}+\frac{1}{4}(Q^\alpha-\tilde Q^\alpha)g_{\mu\nu}-\frac{1}{4}\delta^\alpha_{(\mu}Q_{\nu)}\; ,
\eea
and $L^{\alpha}_{\;\;\;\mu\nu}$ is the disformation tensor:
\bea
L^{\alpha}_{\;\;\;\mu\nu}=\frac{1}{2}Q^{\alpha}_{\;\;\;\mu\nu}-Q_{(\mu\nu)}^{\;\;\;\;\;\;\alpha}\; .
\eea

According to \cite{Zhao:2021zab}, Eq.~\eqref{meteq} can be simplified as:
\bea
f'(Q) G_{\mu\nu}-\frac{1}{2}\left(f(Q)-f'(Q)Q\right)+2f''(Q)P^\alpha_{\;\;\mu\nu}\partial_\alpha Q=T_{\mu\nu}\; ,\label{meteqGmunu}
\eea
where $G_{\mu\nu}$ is the usual Einstein tensor defined by the Levi-Civita connection, and GR can be recovered when $f(Q)=Q$. The tensor $T_{\mu\nu}$ denotes the energy–momentum tensor. We introduce the tensor $\mathcal{M}_{\mu\nu}$ as the left-hand side (lhs) of Eq.~\eqref{meteqGmunu}. Therefore, the equations of motion can be rewritten as:
\bea
\mathcal{M}_{\mu\nu}=T_{\mu\nu}\; .
\eea
For simplicity, we  assume that $T_{\mu\nu}$ takes the form of a perfect fluid:
\bea
T^\mu_{\;\;\nu}=\left(\rho+p\right)U^\mu U_\nu+p g^\mu_{\;\;\nu}\; ,
\eea
where $\rho$ denotes the energy density, $p$ the pressure  and $U_\mu$ the four velocity. If the matter Lagrangian is independent of $\Gamma$, and considering STG, then the equation of motion for the connection reads \cite{Xu:2019sbp, BeltranJimenez:2018vdo}:
\bea
\nabla_\mu\nabla_\nu(\sqrt{-g}f'(Q)P^{\mu\nu}_{\;\;\;\;\;\;\alpha})=0\; ,\label{coneq}
\eea
which we express as $\mathcal{C}_\alpha= 0$ for the sake of notational simplicity.

In this paper, we mainly focus on $f(Q)$ cosmology and its relation with the choice of connections. In the next section, we begin our study by explicitly exploring the set of symmetries for the Friedmann-Lemaître-Robertson-Walker (FLRW) metric  to assure that the components of the connection satisfy them together with the conditions of STG, which are vanishing Riemann and torsion tensors.

\section{FLRW Cosmology}\label{cosmof(Q)}

We will consider $f(Q)$ cosmology and show how the connections play the role in governing the cosmological dynamics of the universe.  The following metric is the FLRW metric which describes the large-scale behavior and phenomenology of our universe \cite{BeltranJimenez:2019tme}:
\begin{equation}
    \dd s^2 = - N(t) \dd t^2 + a^2(t) \left[\dd r^2 + r^2 \left( \dd \theta ^2 + \sin ^2 \theta \dd \phi ^2\right)\right]\; ,
\end{equation}
where $N(t)$ and $a(t)$ are the lapse function and the scale factor as functions of the cosmic time $t$, respectively.

This spacetime is invariant under rotations and spatial translations. This set of symmetries can be described through its Killing vectors \cite{Hohmann:2019nat}:
\begin{align}
\mathcal{R}_x&=\left(0,0,\sin\phi,\frac{\cos\phi}{\tan\theta} \right)& \mathcal{T}_x&=\left(0,\sin\theta\cos\phi,\frac{\cos\theta\cos\phi}{r},-\frac{\sin\phi}{r\sin\theta}\right)\nonumber\\
\mathcal{R}_y&=\left(0,0,-\cos\phi,\frac{\sin\phi}{\tan\theta}\right)&\mathcal{T}_y&=\left(0,\sin\theta\sin\phi,\frac{\cos\theta\cos\phi}{r},\frac{\cos\phi}{r\sin\theta}\right)\nonumber\\
\mathcal{R}_z&=(0,0,0,-1)& \mathcal{T}_z&=\left(0,\cos\theta,-\frac{\sin\theta}{r},0\right)\; ,
\end{align}
where $\mathcal{R}_i$ generate rotations and $\mathcal{T}_i$ generate translations. Consequently, the Lie derivatives of the metric with respect to these killing vector fields vanish:
\bea
\mathcal{L}_{\xi_i} g_{\mu\nu}=0\; ,
\eea
where $\mathcal{L}_{\xi_i}$ is the Lie derivative with respect to the killing vector $\xi_i$. As mentioned in the previous section, these symmetries must be carried over to the connection. Mathematically, this is translated into imposing:
\bea
\mathcal{L}_{\xi_i} \Gamma^\lambda_{\;\;\mu\nu} = 0\; .
\eea

In this way, we demand that the connection respects homogeneity and isotropy. Although we have introduced six Killing vectors, it will be sufficient to use those associated with rotations and one with translations, since invariance under the remaining generators can then be derived from their commutation relations \cite{Hohmann:2019fvf}. This results in 256 equations \cite{Dimakis:2023uib}:
\bea
\mathcal{L}_{\xi_i} \Gamma^{\lambda}_{\mu\nu} = \xi^k_i \frac{\partial \Gamma^{\lambda}_{\mu\nu}}{\partial x^k} 
- \Gamma^{\kappa}_{\mu\nu} \frac{\partial \xi^{\lambda}_i}{\partial x^{\kappa}} 
+ \Gamma^{\lambda}_{\kappa\nu} \frac{\partial \xi^{\kappa}_i}{\partial x^{\mu}} 
+ \Gamma^{\lambda}_{\mu\kappa} \frac{\partial \xi^{\kappa}_i}{\partial x^{\nu}} 
+ \frac{\partial^2 \xi^{\lambda}_i}{\partial x^{\mu} \partial x^{\nu}} = 0\,.
\eea
In addition, we should implement the torsionless ($\Gamma^{\alpha}_{\mu\nu}=\Gamma^{\alpha}_{\nu\mu}$) and flatness ($R(\Gamma)=0$) conditions to built the symmetric flatness background of STG, which leads us to a set of equations for the 64 components of the connection. Then, there are three different and possible choices of the components of the connections as a result of the freedom in solving the equations associated with the symmetries of the connection \cite{DAmbrosio:2021pnd}.

The connection for the case (I), denoted by $\Gamma ^{\text{(I)}}_Q$ reads:
\begin{align}
    &\Gamma ^t {}_{\mu\nu} = \left(\begin{matrix}
        C_3 + \frac{\dot{C}_3}{C_3} & 0 & 0 & 0 \\
        0 & 0 & 0 & 0 \\
        0 & 0 & 0 & 0 \\
        0 & 0 & 0 & 0
    \end{matrix} \right)\,, &\Gamma ^r{}_{\mu\nu} = \left( \begin{matrix}
        0 & C_3 & 0 & 0 \\
        C_3 & 0 & 0 & 0 \\
        0 & 0 & - r & 0 \\
        0 & 0 & 0 & - r \sin ^2 \theta
    \end{matrix} \right)\,, \nonumber \\
    &\Gamma ^\theta {}_{\mu\nu} = \left(\begin{matrix}
        0 & 0 & C_3 & 0 \\
        0 & 0 & \frac{1}{r} & 0 \\
        C_3 & \frac{1}{r} & 0 & 0 \\
        0 & 0 & 0 & -\cos \theta \sin \theta
    \end{matrix} \right)\,, &\Gamma ^\phi{}_{\mu\nu} = \left( \begin{matrix}
        0 & 0 & 0 & C_3 \\
        0 & 0 & 0 & \frac{1}{r}\\
        0 & 0 & 0 & \cot \theta \\
        C_3 & \frac{1}{r} & \cot \theta & 0
    \end{matrix} \right)\,,
\end{align}
where the dot denotes derivatives with respect to $t$, and $C_i$ is a function of time.

The connection for the case (II), denoted by $\Gamma ^{\text{(II)}}_Q$ reads\footnote{ Notice that we have used $\tilde{C}_2$ here in order to distinguish it from the rescaled one $C_2\equiv \tilde{C}_2/a^2$. The latter will be used later in the expression of the field equations to follow the notation used in Ref.~\cite{DAmbrosio:2021pnd}.}:
\begin{align}
    &\Gamma ^t {}_{\mu\nu} = \left(\begin{matrix}
        - \frac{\dot{\tilde{C}}_2}{\tilde{C}_2} & 0 & 0 & 0 \\
        0 & \tilde{C}_2 & 0 & 0 \\
        0 & 0 & \tilde{C}_2 r^2 & 0 \\
        0 & 0 & 0 &\tilde{C}_2 r^2 \sin ^2 \theta
    \end{matrix} \right)\,, &\Gamma ^r{}_{\mu\nu} = \left( \begin{matrix}
        0 & 0 & 0 & 0 \\
        0 & 0 & 0 & 0 \\
        0 & 0 & - r & 0 \\
        0 & 0 & 0 & - r \sin ^2 \theta
    \end{matrix} \right)\,, \nonumber \\
    &\Gamma ^\theta {}_{\mu\nu} = \left(\begin{matrix}
        0 & 0 & 0 & 0 \\
        0 & 0 & \frac{1}{r} & 0 \\
        0 & \frac{1}{r} & 0 & 0 \\
        0 & 0 & 0 & -\cos \theta \sin \theta
    \end{matrix} \right)\,, &\Gamma ^\phi{}_{\mu\nu} = \left( \begin{matrix}
        0 & 0 & 0 & 0 \\
        0 & 0 & 0 & \frac{1}{r}\\
        0 & 0 & 0 & \cot \theta \\
        0 & \frac{1}{r} & \cot \theta & 0
    \end{matrix} \right)\,.\label{connec2component}
\end{align}
Finally, for the connection for the case (III), i.e., $\Gamma ^{\text{(III)}}_Q$, we have:
\begin{align}
    &\Gamma ^t {}_{\mu\nu} = \left(\begin{matrix}
        C_1 & 0 & 0 & 0 \\
        0 & 0 & 0 & 0 \\
        0 & 0 & 0 & 0 \\
        0 & 0 & 0 & 0
    \end{matrix} \right)\,, &\Gamma ^r{}_{\mu\nu} = \left( \begin{matrix}
        0 & 0 & 0 & 0 \\
        0 & 0 & 0 & 0 \\
        0 & 0 & - r & 0 \\
        0 & 0 & 0 & - r \sin ^2 \theta
    \end{matrix} \right)\,, \nonumber \\
    &\Gamma ^\theta {}_{\mu\nu} = \left(\begin{matrix}
        0 & 0 & 0 & 0 \\
        0 & 0 & \frac{1}{r} & 0 \\
        0 & \frac{1}{r} & 0 & 0 \\
        0 & 0 & 0 & -\cos \theta \sin \theta
    \end{matrix} \right)\,, &\Gamma ^\phi{}_{\mu\nu} = \left( \begin{matrix}
        0 & 0 & 0 & 0 \\
        0 & 0 & 0 & \frac{1}{r}\\
        0 & 0 & 0 & \cot \theta \\
        0 & \frac{1}{r} & \cot \theta & 0
    \end{matrix} \right)\,.
\end{align}

Then, there will be three sets of equations of motion for each of the three possible connections. Let us write explicitly the nonzero terms, and recall that $\mathcal{M}_{\mu\nu}=T_{\mu\nu}$, while\footnote{In addition, for each connection, the equations $\mathcal{C}_r=0$, $\mathcal{C}_\theta=0$ and $\mathcal{C}_\phi=0$ are automatically satisfied.} $\mathcal{C}_\alpha=0$ . For the connection $\Gamma ^{\text{(I)}}_Q$, these are:
\begin{align}
    Q &= \frac{3}{2N^2}\left(  C_3 ( 6 H N-\dot{N} )+2 N \dot{C}_3 - 4 H^2 N\right) \,,\label{eq1Q}\\
    \mathcal{M}^t{}_t &= - \frac{3 f' \left( C_3 (\dot{N}-6 HN) -2 N \dot{C}_3 +8 H^2 N\right) + 2 N (3 C_3 f'' \dot{Q} + fN)}{4 N^2} \,,\label{eq1mtt}\\
    \mathcal{M}^r{}_r &= \frac{f' \left(3 C_3 (6HN - \dot{N}) + 6 N \dot{C}_3 - 8 N \dot{H}-24 H^2 N + 4 H \dot{N} \right) - 2N \left( f'' \dot{Q}(4H-3 C_3) + f N\right)}{4N^2} \,,\label{eq1mrr}\\
    \mathcal{C}_t &= - \frac{3 C_3}{4N^2}\left(f'' \dot{Q} (6HN - \dot{N})+2 f''' N \dot{Q}^2 + 2N f'' \ddot{Q}\right)\,.\label{eq1ct}
\end{align}
If $C_3\neq 0$, one can already solve the equation $\mathcal{C}_t$=0 to get:
\bea
3 H f'' \dot{Q}-\frac{1}{2}f'' \dot{Q}\frac{d}{dt}(\ln N)+\frac{d}{dt}(f'' \dot{Q})=0\,,
\eea
which is reduced to $3 H (f'' \dot{Q})+\frac{d}{dt}(f'' \dot{Q})=0$ when we fix $N=1$. Then, we can get:
\bea
f'' \dot{Q}\propto a^{-3}\,.\label{fqda3}
\eea

For the second connection $\Gamma ^{\text{(II)}}_Q$, we have rescaled the function $\tilde{C}_2$ in Eq.~\eqref{connec2component} as $\tilde{C}_2=a^2(t) C_2(t)$, such that now we recover the equations of \cite{DAmbrosio:2021pnd} with the same notation:

\begin{align}
    Q &= \frac{3}{2N}\left( C_2 ( 6 H N+ \dot{N})+2 N \dot{C}_2 - 4 H^2 \right) \,,\label{eq2Q}\\
    \mathcal{M}^t{}_t &= \frac{3 f' \left( C_2 (6 HN + \dot{N}) + 2 N \dot{C}_2 -8 H^2 \right) - 2 N (f  - 3 C_2 f'' \dot{Q})}{4 N} \,,\label{eq2mtt}\\
    \mathcal{M}^r{}_r &= \frac{f' \left(3 C_2 N (6HN + \dot{N}) + 6 N^2 \dot{C}_2 - 8 N \dot{H}-24 H^2 N + 4 H \dot{N} \right) - 2N \left( f'' \dot{Q}(4H- C_2 N) + f N\right)}{4N^2} \,,\label{eq2mrr}\\
    \mathcal{C}_t &=  \frac{3 C_2}{4N}\left(f'' \dot{Q} (10HN + \dot{N})+2 f''' N \dot{Q}^2 + 2N f'' \ddot{Q}\right) + 3 \dot{C}_2 f'' \dot{Q}\,.\label{eq2ct}
\end{align}
In this case,  the equation for $\mathcal{C}_t$ can be written as:
\bea
\frac{d}{dt}(f'' \dot{Q})+5 H f'' \dot{Q}+\frac{1}{2}\frac{\dot{N}}{N}f''\dot{Q}=-2f''\dot{Q}\frac{d}{dt}(\ln C_2)\,.\label{eqconn3c}
\eea
If we assume $N=1$ and define $X\equiv a^5f''\dot{Q}$, we can  simplify Eq.~\eqref{eqconn3c} to get:
\bea
\dot{X}=-2 X\frac{d}{dt}\ln C_2\, ,
\eea
with the solution:
\bea
X\propto C_2^{-2}\;\;\;\;\; \rightarrow \;\;\;\;\; f''\dot{Q}\propto \frac{1}{C_2^{2} a^5}\, .
\eea

Similarly, for the third connection $\Gamma ^{\text{(III)}}_Q$ we obtain:
\begin{align}
    Q &= - \frac{6H^2}{N}\,,\label{eq3Q}\\
    \mathcal{M}^t{}_t &= - \frac{6H^2 f'}{N} - \frac{f}{2}\,,\label{eq3mtt}\\
    \mathcal{M}^r{}_r &= \frac{f' \left(H \dot{N}-2 N \left(\dot{H}+3 H^2\right)\right)-2 H N \dot{Q}
   f''}{N^2}-\frac{f}{2} \,,\label{eq3mrr}\\
    \mathcal{C}_t &=0\,.\label{eq3ct}
\end{align}
One can see that for the connection $\Gamma ^{\text{(III)}}_Q$, the connection degree of freedom does not enter the field equations. Therefore, the cosmological dynamics is, upon choosing a specific lapse $N$, entirely determined by the evolution of the scale factor $a$ and its derivatives. 

We would like to emphasize that all three connections, i.e., $\Gamma ^{\text{(I)}}_Q$, $\Gamma ^{\text{(II)}}_Q$, and $\Gamma ^{\text{(III)}}_Q$, can all be recast to the coincident gauge. However, only the connection $\Gamma ^{\text{(III)}}_Q$ in the coincident gauge allows for the standard diagonal expression for the FLRW metric. The metric expressions for the other two connections, i.e., $\Gamma ^{\text{(I)}}_Q$ and $\Gamma ^{\text{(II)}}_Q$, acquire several off-diagonal components and take non-standard forms in the coincident gauge \cite{Jensko:2024bee}. Therefore, each of them is physically different from the others, as we will show later.

\section{Analytical solutions}\label{subsec:maximsym}

With the equations of motion to be solved being presented, in this section, we will seek analytical solutions under some specific configurations. From now on, we will set the lapse function $N(t)=1$ when discussing the cosmological solutions at the background level. In fact, this lets us write $Q$ for the three connections in a compact and useful form as follows:
\bea
Q_{\Gamma^{(\cdot)}}(t) = 3 \left(3 C_{2,3} H - 2H^2 + \dot{C}_{2,3} \right)\,.\label{genQ}
\eea
where $\Gamma^{(\cdot)}$ denotes one of the three possible connections, taking the corresponding $C_i$ associated with the connection being studied at each moment. Note that the $C_1$ function of $\Gamma ^{\text{(III)}}_Q$ is absent in \eqref{genQ}, recovering the non-metricity scalar \eqref{eq3Q} for the third connection.

\subsection{Maximally symmetric spacetimes}

\subsubsection{Background solutions}

\textbf{Connection $\Gamma^{\text{(I)}}_Q$: }

We start solving the equations of motion for the first connection $\Gamma^{\text{(I)}}_Q$ by considering the maximally symmetric spacetimes with constant $H\equiv\dot{a}/a=H_0$ and in presence of matter with a constant equation of state\footnote{From now on, a subscript 0 stands for constant quantities.} $p(t)=w_0 \rho(t)$, i.e., $\mathcal{M}^t{}_t=-\rho$ and $\mathcal{M}^r{}_r=p$. By subtracting these two equations, i.e., $\mathcal{M}^t{}_t-\mathcal{M}^r{}_r=-\rho-p$, we get:
\bea
(2H_0-3C_3)\dot{Q}f''(Q)=-(1+w_0)\rho\,. \label{c3I}
\eea

\paragraph{Vacuum and cosmological constant case}:
The simplest case corresponds to vacuum, where $\rho=0$ and $p=0$. In principle, there are two possibilities in order to satisfy Eq. \eqref{c3I}:  
\begin{itemize}
    \item Considering that the first term of the lhs of Eq. \eqref{c3I} vanishes, $C_3$ becomes a constant of the form $C_3=2H_0/3$, which, according to Eqs.~\eqref{eq1Q} and \eqref{eq1mtt}, implies $Q=0$ and the function $f(Q)=\beta e^{-\frac{Q}{6H_0^2}}$, respectively. Here, $\beta$ is an arbitrary constant. In addition, $\mathcal{C}_t=0$ will be automatically satisfied since $Q$ becomes a constant.
    \item The other possibility is that $\dot{Q}f''(Q)=0$, without any restriction on $C_3$. Then, either $Q=Q_0$ being $Q_0$ a constant, or $f(Q)=\alpha+\beta Q$, where $\alpha$ and $\beta$ are arbitrary constants, which reduces to GR. If we consider the case with $Q=Q_0$, then $C_3=(Q_0+6 H_0^2)/9H_0+\beta e^{-3H_0 t}$, where $\beta$ is an integration constant. Taking $Q_0$ and $\beta$ to be zero, the previous item is recovered. On the other hand, for $Q_0\neq0$ and $\beta\neq0$, and using \eqref{eq1mrr}, we obtain $f(Q)=(-6 H_0^2+ Q_0)\gamma$, being $\gamma$ an integration constant. Therefore,  we recover GR in this case as well.
\end{itemize}

The above results show that it is possible to have exact solutions of maximally symmetric spacetimes in vacuum for the connection $\Gamma ^{\text{(I)}}_Q$. A similar reasoning emerges when considering a cosmological constant, i.e. $\rho=\rho_0$ and $w_0=-1$, and the rhs of Eq. \eqref{c3I} vanishes. Like in the vacuum case, there are two possible ways. Considering the first one, i.e. $C_3=2 H_0/3$, the solution for this case will be:
\bea
f(Q)=2\rho_0+\beta e^{-\frac{Q}{6 H_0^2}}\,,
\eea
which is the same solution as in the vacuum case with an additional $2\rho_0$ term that is a rescalling of the energy density associated with the cosmological constant. The other possibility, in which $\dot{Q}f''(Q)$ vanishes to satisfy Eq. \eqref{c3I} with $w_0=-1$, also gives the same solutions of the vacuum case with an additional term $2\rho_0$ in the $f(Q)$. 

\paragraph{Matter case}:
We now consider the presence of matter. Substituting the constraint $\mathcal{C}_t = 0$, which implies $\dot{Q}f''(Q) = C_0 / a^3$, where $C_0$ is an arbitrary constant, into Eq.~\eqref{c3I}, we obtain:
\bea
C_3=\frac{2}{3}H_0+\frac{1+w_0}{3C_0}\rho a^3\; .\label{c3sol}
\eea
as long as $C_0\neq 0$.

From Eq.~\eqref{c3sol}, we can see that the case with $\rho\propto a^{-3}$ is a special case in the presence of matter for which $C_3$ becomes constant again. Nevertheless, from the continuity equation, it follows that an energy density evolving in this way corresponds to dust with $w_0=0$. In other words, the dust case forces $C_3$ to be a constant, but this will imply a constant $Q$ due to Eq.~\eqref{eq1Q}. Consequently, $\dot{Q}=0$, and Eq. \eqref{c3I} will imply $\rho=0$. Summarizing, assuming that the matter content is given by dust, only the vacuum scenario in which $\rho=0$ is possible to have maximally symmetric spacetime configurations.

For other cases where $w_0\ne0$ such that the scale factor cannot be simplified, we need to introduce the equation corresponding to the sum $\mathcal{M}^t{}_t+\mathcal{M}^r{}_r$ to obtain an additional relation that close the system. In addition, we can use the expression for $C_3$ given in Eq. \eqref{c3sol}, together with $\dot{a}=H_0a$, to obtain a simpler differential equation for $f(Q)$:
\bea
-\frac{3 H_0  \left(\left(w_0^2-1\right) a^3 \rho+2 C_0
   H_0\right)}{C_0}f'(Q)=\frac{2 C_0 H_0}{a^3}+f(Q)+(w_0-1) \rho \; ,\label{con1matI}
\eea

Analogously, we can introduce the solution of $C_3$ in the expression of $Q(t)$ as well. This allows us to obtain the relation between the non-metricity scalar and the scale factor as follows: 
\bea
Q=-\frac{3 H_0 \left(w_0^2-1\right) a^3 \rho }{C_0}\; .\label{Qa1I}
\eea
Taking into account that the energy density can be expressed as a function of the scale factor via the continuity equation, we can use this relation in Eq. \eqref{con1matI} to obtain a differential equation 
fully written either in terms of $a$ or in terms of $Q$, depending on which is more convenient. In order to clarify this, let us exemplify the radiation case where $w_0=1/3$ and $\rho=\rho_0 a^{-4}$, where $\rho_0$ is a constant that represents the energy density at $a=1$. Then for radiation, Eq. \eqref{con1matI} becomes:
\bea
(6 H^2 - Q) f'(Q)+f(Q) = \frac{27 C_0^4 (Q-8 H^2)\, Q^3}{2048 H^4 \rho_0^3}\,.
\eea
Solving this equation allows us to obtain the $f(Q)$ solution for the radiation case in a maximally symmetric spacetime, given by:
\bea
f(Q)=c_1 \left(Q-6 H_0^2\right)-\frac{9 C_0^4 \left(1296 H_0^8-216 H_0^6
   Q+Q^4\right)}{2048 H_0^4 \rho_0^3}\,.\label{fqradiationdesitter}
\eea
where $c_1$ is a constant of integration. Consequently, we have demonstrated that it is possible to achieve de Sitter-type solutions for the connection $\Gamma^{\text{(I)}}_Q$ with matter - as long as the matter is not dust - with an equation of state $\rho=w_0 p$. This type of solution is interesting, in particular, in its relation to addressing cosmological singularities. For instance, the $f(Q)$ model of Eq.~\eqref{fqradiationdesitter} may shed light on resolving the Big Bang singularity in a radiation-dominated universe by replacing the singularity with an early de Sitter phase, followed by a smooth matter-dominated era, and so on. In fact, non-singular cosmologies with a similar behavior can also be realized in different connections when choosing different functional $f(Q)$, as we will demonstrate in Sec.~\ref{sec:fqcosmology}. 

\vspace{1cm}

\textbf{Connection $\Gamma^{\text{(II)}}_Q$:}

For the connection $\Gamma^{\text{(II)}}_Q$, we will follow the same strategy that we used for the connection $\Gamma^{\text{(I)}}_Q$. Then, the equation $\mathcal{M}^t{}_t-\mathcal{M}^r{}_r=-\rho-p$ reads now:
\bea
(2 H_0+C_2) \dot{Q} f''(Q)+(1+w_0) \rho=0\label{con2-}\,.
\eea
\setcounter{paragraph}{0}
\paragraph{Vacuum and cosmological constant case} :
As before, the vacuum case; i.e.  $\rho = 0$ and $p=0$, is a straightforward case that can be divided into possibilities:
\begin{itemize}
    \item One possibility reads $C_2=-2H_0$, which implies $Q=-24 H_0^2$ from Eq.~\eqref{eq2Q} and $f(Q)=\beta e^{-\frac{Q}{30 H_0^2}}$ from Eq.~\eqref{eq2mtt}.
    \item The other possibility implies $\dot{Q}f''(Q)=0$. Similar to the previous case, there are two options: either $\dot{Q}=0$, or $f(Q)=\alpha+\beta Q$. Both cases correspond to GR. In order to verify this claim for $\dot{Q}=0$, we take  $Q=Q_0$ as a constant. Consequently, we can obtain $C_2=(6 H_0^2 + Q_0)/(9 H_0)+e^{-3H_0 t}\beta$, and from $\mathcal{M}^t{}_t=0$ we finally obtain $f(Q)=(-6H_0^2+ Q)/\gamma$. Thus we see that the current situation is exactly the same as the second case for the connection $\Gamma^{\text{(I)}}_Q$, and in both scenarios, GR is recovered when the integration constants and $Q_0$ are not zero.
\end{itemize}

In the same way as for the connection $\Gamma^{\text{(I)}}_Q$, the case with a cosmological constant can be seen as a slight extension of the vacuum case, as the rhs of Eq. \eqref{con2-} vanishes. Then, if $C_2=-2 H_0$, the solution becomes:
\bea
f(Q)=2\rho_0+\beta e^{-\frac{Q}{30 H_0^2}}\,,
\eea
while the case with $\dot{Q}f''(Q)=0$ implies $f(Q)=2\rho_0-6\beta H_0^2+\beta Q$ for $\dot{Q}=0$, or $f(Q)=\alpha+\beta Q$ for $f''(Q)=0$. Therefore, the case with cosmological constant will also imply GR as in the vacuum case.

\paragraph{Matter case}: Similar to the case of the connection $\Gamma^{\text{(I)}}_Q$, we start using the constraint $\mathcal{C}_t=0$; i.e.  $\dot{Q}f''(Q)=C_0/(C_2^2 a^5)$, in Eq. \eqref{con2-}. This allows us to derive the expressions for $C_2$ as:
\bea
C_2=\frac{-C_0\pm\sqrt{C_0\left(C_0-8 a^5 H_0(1+\omega_0)\rho\right)}}{2 a^5(1+\omega_0)\rho}\,,
\eea
and the non-metricity scalar as:
\bea
Q&=&-6 H_0^2+\frac{9 H_0\left(-C_0\pm\sqrt{C_0 (C_0 - 8 a^5 H_0  (1 + \omega_0)\rho}\right)}{2 a^5(1+\omega_0)\rho}\nonumber\\
&-&\frac{3 C_0 H (3 \omega_0-2) \left(\sqrt{C_0\left(C_0-8 H (1+\omega_0) a^5 \rho \right)}\pm 4 H (1+\omega_0) a^5 \rho \mp C_0\right)}{2 (1+\omega_0) a^5 \rho \sqrt{C_0 \left(C_0-8 H (1+\omega_0) a^5 \rho \right)}}\,.
\label{Q2mat}
\eea

In addition, and as done previously, we can use the sum $\mathcal{M}^t{}_t+\mathcal{M}^r{}_r$ to obtain an additional differential equation to close the system:
\bea
3 \left(\dot{C_2}+3 H C_2-4 H_0^2\right) f'(Q)+2 (C_2-H_0) \dot{Q}
   f''(Q)-f(Q)=(w_0-1) \rho\,.  \label{con2+}
\eea

However, because for this case $\dot{Q}f''(Q)\propto1/(C_2^2 a^5)$, considering dust will not imply vacuum. In fact, for  $w_0=0$, the energy density scales as $\rho=\rho_0 a^{-3}$, and $C_2$ is not a constant due to Eq. \eqref{con2+}. Consequently, $C_0$ will not vanish, and it is possible to find dust solutions unlike the result for the connection $\Gamma^{\text{(I)}}_Q$. 

For a general case, one can find solutions following the same reasoning as for the connection $\Gamma^{\text{(I)}}_Q$. First, one should write Eq. \eqref{con2+} either in terms of $a$ or in terms of $Q$, from the continuity equation and the relation between $Q$ and $a$ from Eq. \eqref{Q2mat}. Then, one is able to solve the differential equation to find $f(Q)$. The complication of this case lies in the expression for $C_2$, which, in addition to being more complicated than $C_3$, has two choices as per the $\pm$ signs of the square root term. Consequently, the expressions for $f(Q)$ obtained for dust and radiation will also be complicated.

\vspace{1cm}

\textbf{Connection $\Gamma^{\text{(III)}}_Q$:}

With the energy density $\rho$ and pressure $p=w_0\rho$, the equation $\mathcal{M}^t{}_t-\mathcal{M}^r{}_r=-\rho-p$ for the connection $\Gamma_Q^{\text{(III)}}$ can be written as: 
\bea
2 H_0 \dot{Q}f''(Q)=-\rho(1+w_0)\,.\label{mss3eq}
\eea
However, since $Q=-6 H^2_0$, then $\dot{Q}=0$ and consequently the solution must be vacuum ($\rho=0$) or a cosmological constant with $p = - \rho$ if $f''(Q)$ is finite.

For the vacuum case and using $\mathcal{M}^t{}_t=0$, one obtains:
\bea \label{f(Q)-conIII-vacuum}
f(Q)=\beta e^{-\frac{Q}{12 H_0^2}}\,,
\eea
but for the cosmological constant case with $\mathcal{M}^t{}_t=-\rho_0$:
\bea
f(Q)=2\rho_0+\beta e^{-\frac{Q}{12 H_0^2}}\,.
\eea

As a result, the connection $\Gamma_Q^{\text{(III)}}$ imposes significantly stronger constraints on de Sitter solutions, since $Q$ must be constant. This constraint arises because the free parameter linked to the connection, $C_1$, does not enter the equations, leaving the system's dynamics entirely governed by the Hubble function $H$.  However, in a maximally symmetric spacetime, where $H$ is not dynamic, the other components of the system are also non-dynamic. Consequently, the energy density $\rho$ must also be constant. If this constant equals zero, the solution corresponds to a vacuum; otherwise, it represents a cosmological constant.

\subsubsection{Stability of the vacuum solutions analysis}

Here, we examine the linear stability of the system by performing small perturbations around the background spacetime, and see whether these perturbations grow or decay over time, indicating the presence of instability or stability, respectively. The full ansatz including the background spacetime with subscript 0 and tiny perturbations with subscript $p$ can be written as
\begin{equation}\label{perturbations}
    H(t) = H_0 + H_p(t) \,,\quad N(t) = 1 + N_p(t) \,,\quad C_i (t) = C_{i,0} + C_{i,p}(t)\,,
\end{equation}
where the subscript $i$ represents three different connections. The substitution of above ansatz into the equations of motion yields differential equations for the perturbation functions at the first-order perturbation, which can be solved analytically for three connections as follows.

\textbf{Connection $\Gamma^{\text{(I)}}_Q$:}
The background, i.e. $0$th order equations are identically satisfied, while the $1$st order ones will give,
\begin{gather}
    H_p(t) = c_1 + \frac{H_0}{2} N_p(t) \,,\\
    C_{3,p}(t) = (c_2 + c_3 t )e^{-3 H_0 t} + 2 c_1 + \frac{H_0}{3} N_p(t)\,,
\end{gather}
for any integration constant $c_1, c_2, c_3$ and arbitrary $N_p(t)$. The non-metricity scalar takes the form 
\bea
Q = 3  e^{-3 H_0  t} \left(c_3+4 c_1 H_0  e^{3 H_0  t}\right)\,.
\eea
Since the system is satisfied for any $N_p(t)$, we can set it to zero without loss of generality. This means that, at least at linear order, our solution is stable for $H_0>0$, as small perturbations introduced to the system decay over time rather than amplifying. 

\textbf{Connection $\Gamma^{\text{(II)}}_Q$:}
As expected, the $0$th order equations of the set \eqref{eq2mtt}-\eqref{eq2ct} are satisfied, while the $1$st order ones give
\begin{equation}
    H_p(t) = c_1  \,,\quad     C_{2,p}(t) = \frac{14 c_1}{3} - \frac{e^{-5 H_0  t} \left(3 c_2 + 5 c_3 e^{2 H_0  t}\right)}{15 H_0 }\,,
\end{equation}
for $c_1, c_2$ and $c_3$ being integration constants and the non-metricity scalar (background plus 1st order) is given by
\begin{equation}
    Q = -24 H_0 ^2 + \frac{6 c_2}{5}  e^{-5 H_0  t} + 12 c_1 H_0 \,.
\end{equation}
Again, we see that at linear order for $H_0>0$ the perturbations decay with time. In the above we set $N_p(t) = 0$ since it does not affect the results.

\textbf{Connection $\Gamma^{\text{(III)}}_Q$:} Here, Eq. \eqref{perturbations} takes the form
\begin{equation}
    H(t) = H_0 + \frac{1}{2}H_0 N_p(t) \,,\quad N(t) = 1 + N_p(t) \,,\quad C_{1,p} (t) = 0\,,
\end{equation}
and the $1$st order equations (namely Eq. \eqref{eq3mtt} - \eqref{eq3ct}) are satisfied for any $N_p(t)$. The non-metricity scalar is unaffected from the perturbations at linear order, i.e. $Q = - 6 H_0^2$, ensuring stability.

\subsection{Vacuum solutions in power law $f(Q)$ for different connections}
In the previous subsection, we demonstrated how different choices of connections in $f(Q)$ theory can realize maximally symmetric spacetimes in different manners. From a complementary perspective, in the following two subsections, we will consider some specific functional forms of $f(Q)$ and demonstrate their vacuum solutions for each connection.

Let us assume that $f(Q) = Q^\kappa$ and set the lapse function to unity, $N = 1$. In vacuum, the set of equations for both $\Gamma ^{\text{(I)}}_Q$ and $\Gamma ^{\text{(II)}}_Q$, i.e. $\mathcal{M}^t{}_t=0$ and $\mathcal{M}^r{}_r=0$,  are proportional to $Q^{\kappa -2}$, and $\mathcal{C}_t=0$ is proportional to $Q^{\kappa-3}$, where $Q(t)$ is given by Eq. \eqref{genQ}. Thus, as long as $\kappa >3$, it is possible to solve the system by simply solving the equation  $Q(t)=0$. Since $C_2$ or $C_3$ can be chosen freely, this does not affect the scale factor at all. For any scale factor, $a$, one can calculate the corresponding $C_{2,3}$ as:
\begin{equation}
    C_{2,3} = \frac{l_0}{a^3} + \frac{2}{a^3}\int ^t a(\tau) \dot{a}(\tau)^2 \dd \tau\,,\label{C23}
\end{equation}
with $l_0$ being an arbitrary integration constant. Consequently, one can impose a specific form of the scale factor and reconstruct the required $C_{2,3}$ from it. Let us take the example proposed in \cite{DAmbrosio:2021pnd}, where $a=a_0 t^\lambda$, and the solution of Eq. \eqref{C23} becomes:
\bea
C_{2,3}=\frac{l_0}{a_0^3}t^{-3\lambda}-\frac{2\lambda^2}{(1-3\lambda)t}\; .
\eea

However, it is possible to go the opposite way. That is, given a specific form of $C_{2,3}$, one can calculate the corresponding scale factor. Again, if we take the example of \cite{DAmbrosio:2021pnd}, with:
\begin{equation}
    C_{2,3} =  \frac{\sqrt{2\Lambda}}{3}\,,
\end{equation}
we recover the two possible solutions of the scale factor:
\bea
a(t)=a_0 \;\;\;\;\;\; \text{or} \;\;\;\;\;\; a(t)=a_0 e^{\sqrt{\frac{\Lambda}{2}}t}\,.
\eea

Interestingly, for connection $\Gamma ^{\text{(III)}}_Q$, the only possible solutions in vacuum imply $H=0$, exactly as in GR. This solution is essentially $Q(t)=0$, like for the other connections. Therefore, mathematically, it is not a very different solution, although physically and cosmologically it is a (phenomenologically) different solution from those previously found for connections $\Gamma ^{\text{(I)}}_Q$ and $\Gamma ^{\text{(II)}}_Q$.

\subsection{Vacuum solutions in exponential $f(Q)$ for different connections}

Let us now assume that $f(Q) = e^{\kappa Q} + \lambda Q$ and again set the lapse function to unity, i.e., $N = 1$. The non-metricity scalar for both connection $\Gamma ^{\text{(I)}}_Q$ and connection $\Gamma ^{\text{(II)}}_Q$, written in Eq. \eqref{genQ}, which for $Q = 0$ yields the equation \eqref{C23}. Even though both the connection equations, Eq. \eqref{eq1ct} and Eq. \eqref{eq2ct}, are satisfied identically, the equations of $\mathcal{M}^t{}_t$ and $\mathcal{M}^r{}_r$ for both connections are satisfied only for
\begin{equation}
    H(t) = \pm \frac{1}{\sqrt{-6(\kappa+\lambda)}}\,.
\end{equation}
Regarding the third connection, the only possible solution is for $\lambda = 0$ and is 
\begin{equation}
    H^2(t) = - \frac{1}{12 \kappa}\,,
\end{equation}
verifying Eq.\eqref{f(Q)-conIII-vacuum}.

\section{Numerical solutions: Born-Infeld $f(Q)$ cosmology}\label{sec:fqcosmology}
In this section, we provide a more explicit example of how different choices of connections in $f(Q)$ gravity can lead to distinctive behaviors of cosmological solutions. In particular, we will first focus on the cosmological solution in the very early universe, in which the universe is dominated by radiation. Then, we will model the late-time universe by considering a simple phantom dark energy. In GR, the radiation-dominated universe starts with a Big Bang singularity, while a phantom-dominated universe ends up with a Big Rip singularity in the future. In $f(Q)$ gravity, the choices of different connections, on top of a given functional form of $f(Q)$, can alter the behaviors of the cosmological solutions in particular at the regimes near the singularities, or even remove the singularities{\footnote{For instance, the Big Bang singularity in a radiation-dominated universe may be alleviated by considering the model in Eq.~\eqref{fqradiationdesitter} for the connection $\Gamma_Q^{\text{(I)}}$.}}. 

To be more specific, we consider the Born-Infeld functional form
\be
f(Q)=\lambda\left(\sqrt{1+\frac{2Q}{\lambda}}-1\right)\,,\label{BIf}
\ee
where $\lambda$ is the Born-Infeld constant. When $Q/\lambda\rightarrow0$, we have
\be
f(Q)\approx Q-\frac{Q^2}{2\lambda}\,,
\ee
and the theory reduces to GR. However, the square root structure in Eq.~\eqref{BIf} implies that the non-metricity scalar is bounded, which may regularize the cosmological singularities, depending on the choices of connections. We will show later that this is indeed the case for the connection $\Gamma_Q^{\text{(III)}}$ and a positive $\lambda$. We will thus assume $\lambda>0$ from now on.

\subsection{Radiation-dominated universe}

We will first consider a radiation-dominated universe and exhibit how different connections in the Born-Infeld $f(Q)$ gravity alter the cosmological behaviors near the Big Bang singularity. A similar analysis has been carried out in the context of Born-Infeld $f(T)$ gravity \cite{Ferraro:2006jd,Ferraro:2008ey}, and it was shown there that the Big Bang singularity is replaced with an early de Sitter phase due to the fact that the torsion scalar $T$, which is proportional to $H^2$ for flat FLRW cosmology, is bounded by the Born-Infeld action. For $f(Q)$ gravity, it is known that the field equations for the connection $\Gamma_Q^{\text{(III)}}$ have a similar structure to that of flat FLRW spacetime in $f(T)$ gravity \cite{DAmbrosio:2021pnd} and the cosmological solutions are expected to share similar behaviors as well. Therefore, as opposed to the previous sections, here we will first consider the connection $\Gamma_Q^{\text{(III)}}$, and show that the Big Bang singularity can indeed be avoided in a similar manner as that in Born-Infeld $f(T)$ gravity. Then, we will consider the other two connections, and demonstrate how different choices of connections can affect the cosmological dynamics for a given function $f(Q)$.

\subsubsection{Connection \texorpdfstring{$\Gamma_Q^{\text{(III)}}$}{GammaQ(III)}}
\begin{figure}[t]
\centering
\includegraphics[width=250pt]{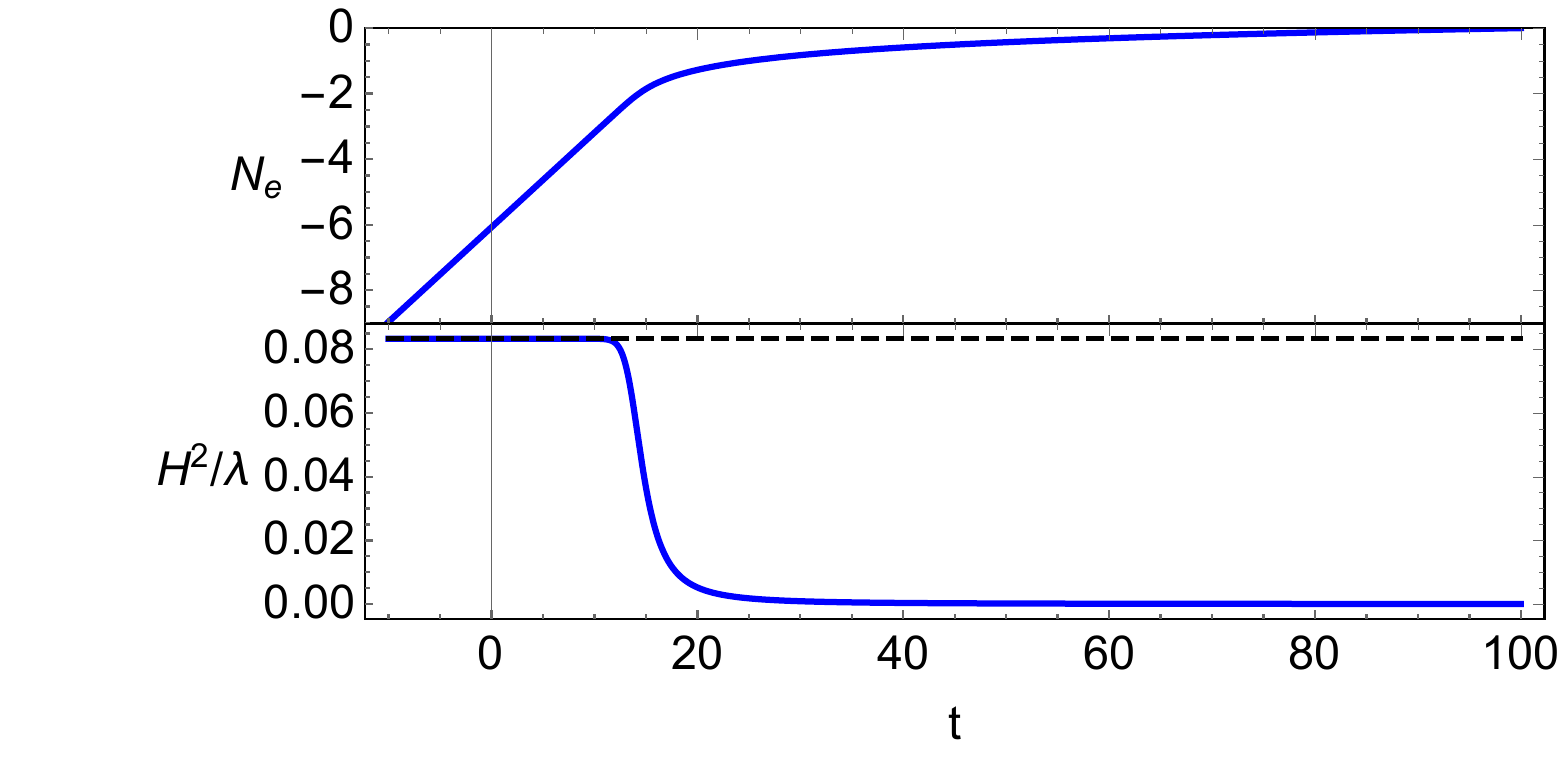}
\includegraphics[width=220pt]{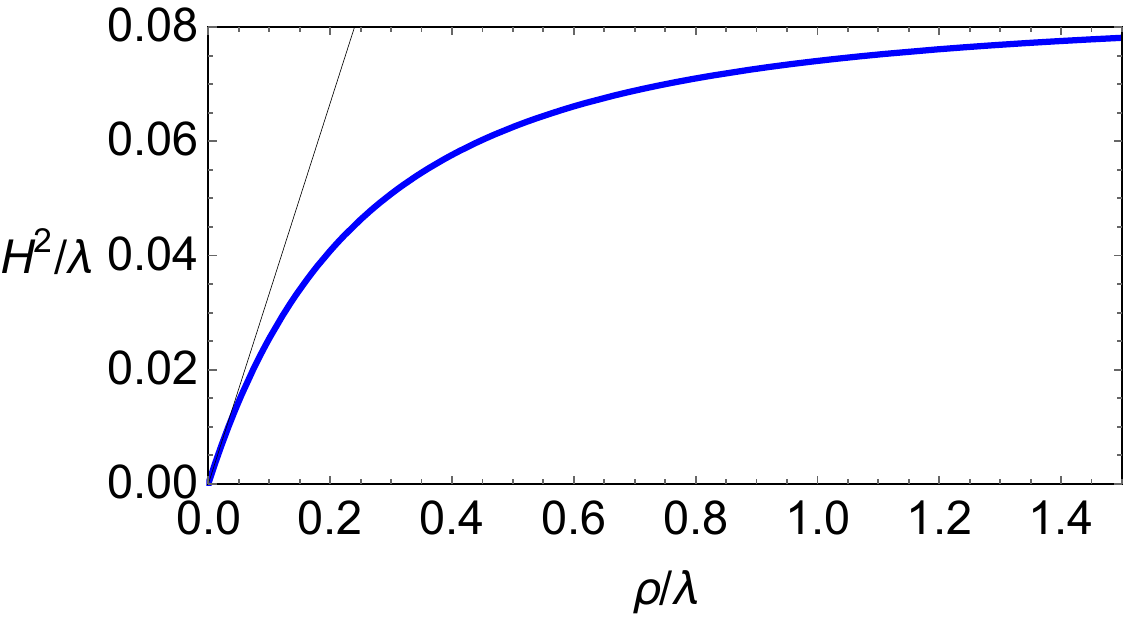}
\caption{(Left) The e-folds $N_e(t)$ and the square of the Hubble function $H^2$ as a function of cosmic time $t$ for the non-singular cosmology for the connection $\Gamma_Q^{\text{(III)}}$. The Hubble function is bounded by $H^2\le\lambda/12$ as shown by the dashed line. (Right) $H^2$ as a function of $\rho$. The standard GR solution $H^2=\rho/3$ is represented by the gray line on the right panel.}
\label{fig:coin}
\end{figure}

We consider the Born-Infeld $f(Q)$ of the form given by Eq.~\eqref{BIf} and start with the connection $\Gamma_Q^{\text{(III)}}$. The field equations for the connection $\Gamma_Q^{\text{(III)}}$, i.e., Eqs.~\eqref{eq3Q}-\eqref{eq3ct}, can be easily integrated. We assume a radiation-dominated universe with $p=w_0\rho$ and $w_0=1/3$. We define the e-fold $N_e\equiv\ln{a/a_0}$ and impose the initial conditions at a late time at $t_0=100$ with $\rho_0/\lambda=10^{-4}$, where $a_0$ and $\rho_0$ are initial values of the scale factor and the energy density at $t_0$, respectively. The initial condition for the Hubble function is determined by the $\mathcal{M}^t{}_t$ equation. We then numerically integrate the equations backward in $t$ and present the results in Fig.~\ref{fig:coin}. As one can see, the Big Bang singularity is replaced by a primordial de Sitter phase in which the Hubble function asymptotically approaches $H\rightarrow\sqrt{\lambda/12}$ when $t\rightarrow-\infty$. Such an early de Sitter phase may be an alternative to inflation and has been identified in Born-Infeld $f(T)$ gravity \cite{Ferraro:2006jd,Ferraro:2008ey}, which shares similar field equations as the case here.

It should be emphasized that the solutions with a de Sitter phase in a radiation dominated universe found here do not contradict the results we obtained in sec.~\ref{subsec:maximsym}. In that section, we showed that for connection $\Gamma_Q^{\text{(III)}}$ with generic $f(Q)$ maximally symmetric solutions are only possible either in vacuum or having a cosmological constant according to Eq.~\eqref{mss3eq}. The reason why we can bypass the argument in sec.~\ref{subsec:maximsym} is because for the de Sitter phase found here, we have
\begin{align}
\dot{Q}&\approx-2\dot{\rho}\left(\frac{\lambda}{2\rho}\right)^3\,,\nonumber\\
f''(Q)&=-\frac{1}{\lambda\left(1+\frac{2Q}{\lambda}\right)^{3/2}}\approx-\frac{1}{\lambda}\left(\frac{2\rho}{\lambda}\right)^3\,,\nonumber
\end{align}
such that Eq.~\eqref{mss3eq} is still satisfied. Therefore, by having some specific $f(Q)$ for which $f''(Q)$ diverges at proper limits, one can still construct solutions with de Sitter phase in the presence of matter fields that are not cosmological constant. This argument is generic in the sense that it applies not only to a radiation-dominated universe, but also to other matter fields, as we will show in sec.~\ref{subsubsec:coinbigrip}.

\subsubsection{Connection $\Gamma ^{\text{(I)}}_Q$}
When choosing the $\Gamma ^{\text{(I)}}_Q$, the connection degree of freedom $C_3$ appears and can affect the non-singular behavior of the solutions we found for the connection $\Gamma ^{\text{(III)}}_Q$. For convenience, we rewrite the expression of the non-metricity \eqref{eq1Q} and the first Friedmann equation \eqref{eq1mtt} as (we set $N=1$)
\begin{align}
Q&=9HC_3+3\dot{C}_3-6H^2\,,\label{Q91}\\
\rho&=3f'H^2-\frac{f'}{2}Q+\frac{f}{2}+\frac{3}{2}C_3f''\dot{Q}\,,\label{rho92}
\end{align}
where the original $\dot{C}_3$ term in the Friedmann equation is replaced using Eq.~\eqref{Q91}.

The first exact solution one can identify is the GR solution with a constant $Q=Q_0$. In this case, the last term on the rhs of Eq.~\eqref{rho92} vanishes. The second and third terms are constant and can be collected as an effective cosmological constant, which does not contribute much at the early time. Eq.~\eqref{Q91} is fulfilled with both $C_3$ and $H$ being dynamical. In this case, we have a Big Bang singularity at $a=0$ where $H^2\approx\rho/3f'$. Note that such GR solution with constant $Q$, irrespective of whatever $f(Q)$ is, has been pointed out in \cite{DAmbrosio:2021pnd}. 

To obtain non-GR solutions, we choose the same initial conditions for $\rho$ and $H$ at $t_0=100$ as those in the previous case, i.e., the connection $\Gamma ^{\text{(III)}}_Q$, from which we obtained non-singular cosmology. The initial conditions for $Q$ and $\dot{Q}$ are determined by those for $C_3$ and $\dot{C}_3$ via Eqs.~\eqref{Q91} and \eqref{rho92}. Therefore, we only have to insert the initial conditions for $C_3$ and $\dot{C}_3$. We then choose the initial condition for $\dot{C}_3$ such that the effective equation of state $w_\textrm{eff}\equiv-2\dot{H}/3H^2-1$ is close to $1/3$ at $t_0$. We then numerically integrate the field equations \eqref{eq1Q}-\eqref{eq1ct} backward in $t$ to obtain solutions. 

\begin{figure}[t]
\centering
\includegraphics[width=300pt]{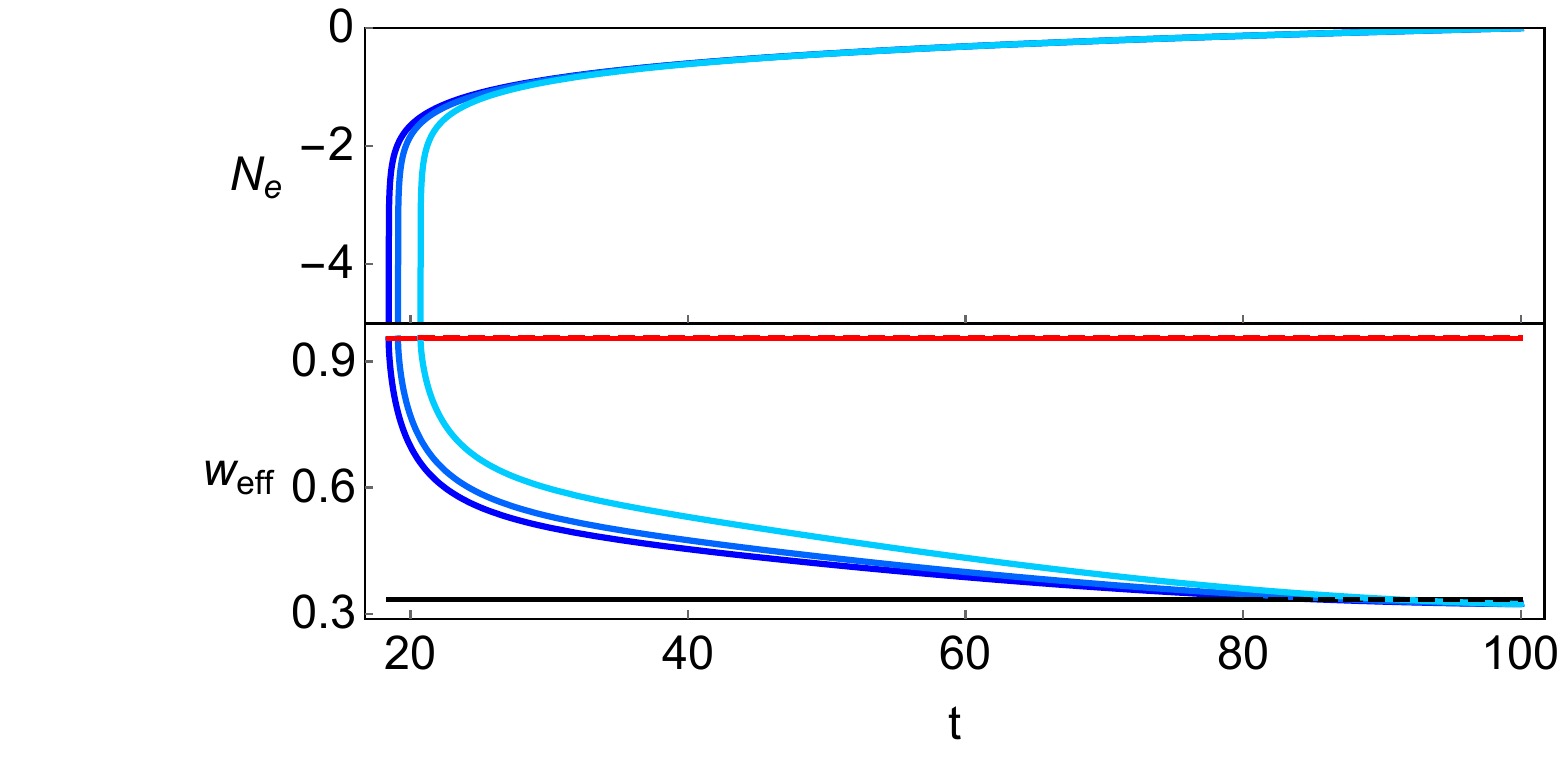}
\caption{The e-folds $N_e(t)$ (top), and the effective equation of state $w_{\textrm{eff}}$ as a function of cosmic time $t$ (bottom) for the cosmology with connection $\Gamma ^{\text{(I)}}_Q$ in a radiation-dominated universe. The initial values for $C_3$ are chosen as $C_3(t_0)=H(t_0)/20$, $H(t_0)/10$, and $H(t_0)/5$, from blue to cyan. The universe has a Big Bang singularity with an effective equation of state close to $w_{\textrm{eff}}\approx 1 $ (red lines in the bottom panel).}
\label{fig:con1}
\end{figure}

We show the numerical results of non-GR solutions in Fig.~\ref{fig:con1}. The initial values for $C_3$ are chosen as $C_3(t_0)=H(t_0)/20$, $H(t_0)/10$, and $H(t_0)/5$, from blue to cyan. The scale factor approaches zero ($N_e\rightarrow-\infty$) at a finite time in the past, corresponding to a Big Bang singularity. Interestingly, we find that the effective equation of state near the singularity approaches that of a stiff matter $w_\textrm{eff}\approx 1-\epsilon$, with $\epsilon$ a tiny value depending on the initial conditions. We can estimate the value of $\epsilon$ (or $w_\textrm{eff}$ when $a=0$) as follows. According to our numerical results, the ratio $C_3/H$ approaches a non-zero finite value near the singularity, i.e., $C_3/H\rightarrow A$, when $a\rightarrow 0$. The value of $A$ is about $\mathcal{O}(10)$ in our numerical results, and it varies with respect to the choice of initial conditions. Also, the non-metricity $Q$ is finite near the singularity, with its time derivative diverging. Therefore, from Eq.~\eqref{Q91}, we find that, at the leading order, $\dot{H}/H^2\rightarrow-3+2/A$ near the singularity, which implies that $w_\textrm{eff}\rightarrow 1-4/(3A)$, as indicated by the red lines in the bottom panel of Fig.~\ref{fig:con1}. 

Note that we have also considered the initial conditions with $C_3(t_0)<0$, and the solutions have similar Big Bang behaviors, thus we do not repeat the results here.

\subsubsection{Connection $\Gamma ^{\text{(II)}}_Q$}

\begin{figure}[t]
\centering
\includegraphics[width=300pt]{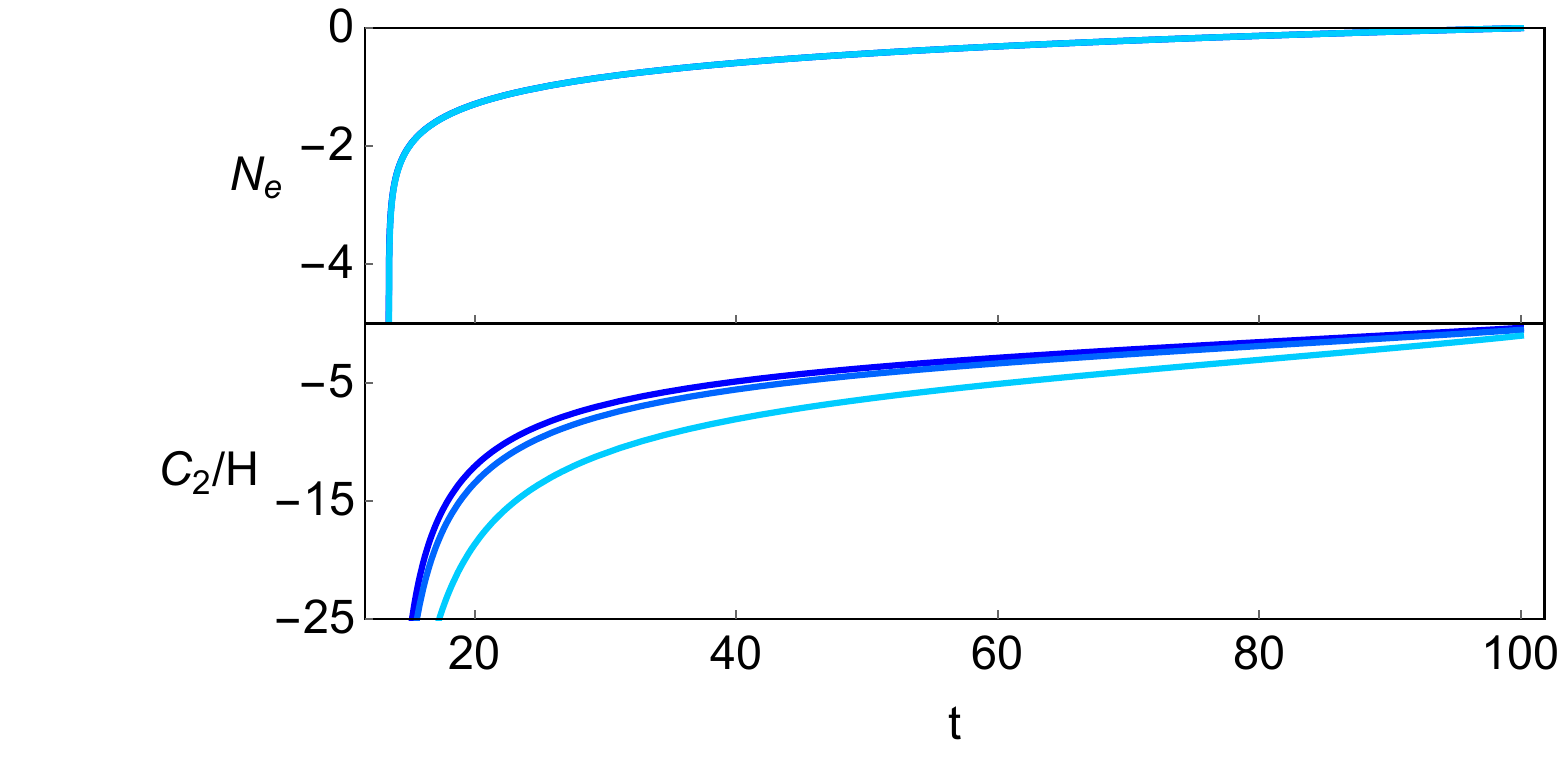}
\caption{The e-folds $N_e$ (top) and the ratio $C_2/H$ (bottom) as functions of cosmic time $t$ for the cosmology with connection $\Gamma ^{\text{(II)}}_Q$. Near the Big Bang singularity where $a\rightarrow0$, we have $H^2a^4$ approaches a non-zero constant, which implies $H^2\propto\rho$. Also, the ratio $C_2/H$ scales as $1/a$ near the singularity. In this figure, the initial conditions for $C_2(t_0)/H(t_0)$ is set to be $-1/3$, $-1/2$, and $-1$, from blue to cyan.}
\label{fig:con2}
\end{figure}

For the connection $\Gamma ^{\text{(II)}}_Q$, the non-metricity \eqref{eq2Q} and the first Friedmann equation \eqref{eq2mtt} can be expressed as
\begin{align}
Q&=9HC_2+3\dot{C}_2-6H^2\,,\label{qeqconnect2}\\
\rho&=3f'H^2-\frac{f'}{2}Q+\frac{f}{2}-\frac{3}{2}C_2f''\dot{Q}\,.\label{rhoconnec2}
\end{align}
In order to find non-GR solutions, we follow a similar method as we adopted for the connection $\Gamma ^{\text{(I)}}_Q$. We choose the same initial conditions for $\rho$ and $H$ at $t_0=100$ as those in the previous cases. The initial conditions for $Q$ and $\dot{Q}$ are determined by those of $C_2$ and $\dot{C}_2$ via Eqs.~\eqref{qeqconnect2} and \eqref{rhoconnec2}. Then, the initial condition for $\dot{C}_2$ is set by assuming that $w_\textrm{eff}$ is close to $1/3$ at $t_0$. We then numerically integrate Eqs.~\eqref{eq2Q}-\eqref{eq2ct} backward in $t$. The results are shown in Fig.~\ref{fig:con2}, in which $C_2(t_0)/H(t_0)$ is set to be $-1/3$, $-1/2$, and $-1$, from blue to cyan. The solutions possess a Big Bang singularity in the past. However, as opposed to the case of connection $\Gamma ^{\text{(I)}}_Q$ where $\dot{Q}$ diverges at the singularity, the value of $Q$ converges to a non-zero finite value toward the past and its derivative $|\dot{Q}|\rightarrow0$ there. Therefore, the solutions approach GR-like solutions near the Big Bang singularity in the sense that the effective equation of state approaches $1/3$, and $H^2a^4$ approaches a non-zero finite value. 

We can also estimate the approximated behavior of $C_2$ near the singularity. Because $Q$ approaches a constant, and the ratio $C_2/H$ diverges (see the bottom panel of Fig.~\ref{fig:con2}), the first two terms on the rhs of Eq.~\eqref{qeqconnect2} balance at the leading order. Therefore, we have $C_2\propto1/a^3$ and $C_2/H\propto1/a$ near the singularity.

\subsection{Phantom-dominated universe}
In this subsection, we assume that the universe is dominated by a phantom dark energy with a constant equation of state $w_0<-1$. In GR, such a phantom-dominated universe will end up with a Big Rip singularity in a finite cosmic time in the future, where the scale factor $a$, the Hubble function $H$, and its cosmic time derivatives diverge. In this subsection, we will show that the phantom-dominated universe in the Born-Infeld $f(Q)$ gravity can have different behaviors, depending on the choices of the connections. Similar to what we have done in the previous subsection, we will start with the connection $\Gamma_Q^{\text{(III)}}$ for which the field equations share the same structure as those of $f(T)$ cosmology. In this case, the Hubble function $H$ is bounded by the Born-Infeld structure, and one also expects that the Big Rip singularity can be avoided. After considering the connection $\Gamma_Q^{\text{(III)}}$, we will choose the other two connections and investigate how the cosmological solutions can be altered. 

\subsubsection{Connection $\Gamma_Q^{\text{(III)}}$}\label{subsubsec:coinbigrip}

\begin{figure}[t]
\centering
\includegraphics[width=300pt]{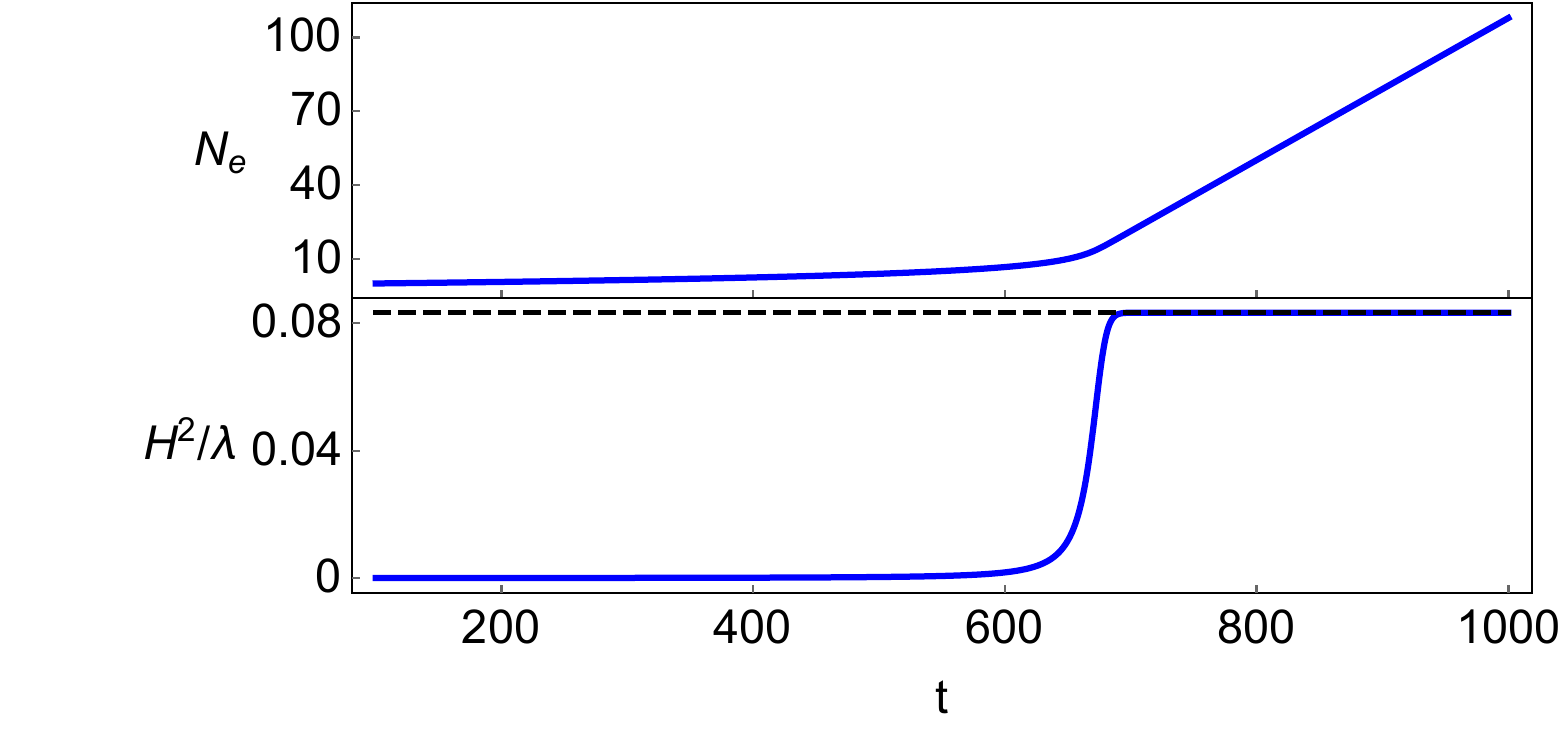}
\caption{The e-folds $N_e(t)$ (top) and the square of the Hubble function $H^2$ as a function of cosmic time (bottom) of the non-singular cosmology for the connection $\Gamma ^{\text{(III)}}_Q$ with a phantom-dominated universe. The Hubble function is bounded by $H^2\le\lambda/12$ as shown by the dashed line in the bottom panel. The Big Rip singularity is replaced by a late-time de Sitter phase. }
\label{fig:coinbigrip}
\end{figure}

For the connection $\Gamma_Q^{\text{(III)}}$, we integrate the field equations~\eqref{eq3Q}-\eqref{eq3ct} assuming a constant equation of state $w_0=-1.2$. The initial conditions are again set at $t_0=100$ and we integrate the equations forward with $\rho_0/\lambda=10^{-4}$. The initial condition for $H$ is determined by the $\mathcal{M}^t{}_t$ equation. The results are shown in Fig.~\ref{fig:coinbigrip}. One can see that the Big Rip singularity is replaced by a de Sitter phase in the future, where the Hubble function reaches its maximum $\sqrt{\lambda/12}$, which is indicated by the dashed line in the bottom panel of Fig.~\ref{fig:coinbigrip}.

\subsubsection{Connection $\Gamma ^{\text{(I)}}_Q$}

\begin{figure}[t]
\centering
\includegraphics[width=300pt]{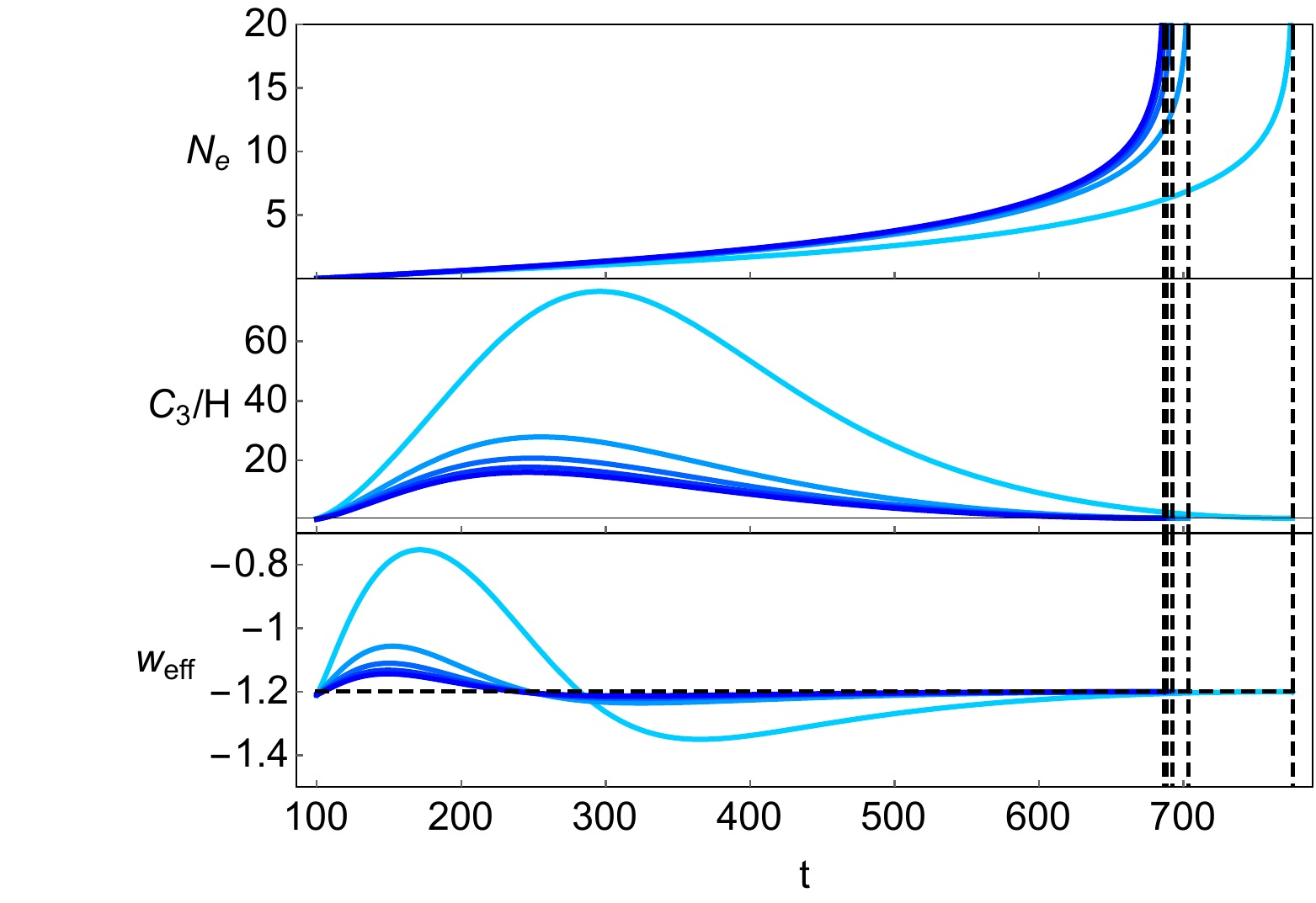}
\caption{The e-folds $N_e(t)$ (top), the ratio $C_3/H$ (middle), and the effective equation of state $w_{\textrm{eff}}$ (bottom) as a function of cosmic time $t$ for the phantom-dominated cosmology with connection $\Gamma ^{\text{(I)}}_Q$. The vertical dashed lines represent the time of the Big Rip singularity. The initial conditions for $C_3$ are chosen as $H(t_0)/C_3(t_0)=4$, $3.5$, $3$, $2.5$, and $2$, from blue to cyan.}
\label{fig:con1bigrip}
\end{figure}

To derive the cosmological solutions for connection $\Gamma ^{\text{(I)}}_Q$, we numerically integrate the field equations \eqref{eq1Q}-\eqref{eq1ct} forward in $t$, with initial conditions for $\rho$ and $H$ at $t_0=100$ identical to those set in sec.~\ref{subsubsec:coinbigrip}. Following a similar method as we did in the case of a radiation-dominated universe, the initial conditions for $Q$ and $\dot{Q}$ are determined by those for $C_3$ and $\dot{C}_3$ via Eqs.~\eqref{Q91} and \eqref{rho92}. We then choose the initial conditions for $\dot{C}_3$ such that the effective equation of state is close to $w_\textrm{eff}\approx-1.2$ at $t=t_0$. Finally, the initial conditions for $C_3$ are chosen as $H(t_0)/C_3(t_0)=4$, $3.5$, $3$, $2.5$, and $2$ with results shown in Fig.~\ref{fig:con1bigrip} ($H(t_0)/C_3(t_0)$ decreases from blue to cyan).

According to our numerical results, the non-metricity $Q$ converges to a non-zero constant when $t$ grows. Therefore, the solutions approach GR-like solutions, ending up with a Big Rip singularity in a finite future $t$ (the vertical dashed lines) where $w_\textrm{eff}\rightarrow w_0$. In particular, near the singularity, the ratio $C_3/H$ converges to 
\be
\frac{C_3}{H}\rightarrow\frac{4}{3(1-w_0)}\,,\label{c3hfinal1}
\ee
as indicated by the horizontal line in the middle panel of Fig.~\ref{fig:con1bigrip}. Note that Eq.~\eqref{c3hfinal1} can be analytically derived by using Eq.~\eqref{Q91}. More explicitly, since $Q$ is finite, and both $H$ and $C_3$ diverge at the same rate, the leading order of the rhs of Eq.~\eqref{Q91} vanishes. Combining with the definition $w_\textrm{eff}\equiv-2\dot{H}/3H^2-1$, one gets Eq.~\eqref{c3hfinal1}.

We would like to mention that although the numerical solutions in this case converge to GR-like solutions when $t$ grows, the effective equation of state $w_\textrm{eff}$ (bottom) has an interesting oscillating behavior in time. The larger $C_3(t_0)/H(t_0)$ is, the more drastic the oscillation pattern in $w_\textrm{eff}$ is developed, and the Big Rip singularity is delayed more. In addition, due to the oscillation pattern of $w_\textrm{eff}$ in $t$, the cosmological solutions may cross the phantom divide line at certain periods during cosmic evolution, as one can see from the cyan line of Fig.~\ref{fig:con1bigrip}. Such scenarios are extremely relevant in relaxing the Hubble tension \cite{DiValentino:2025sru}.

\subsubsection{Connection $\Gamma ^{\text{(II)}}_Q$}
For the connection $\Gamma ^{\text{(II)}}_Q$, we again choose the initial conditions for $\rho$ and $H$ at $t_0=100$. Then, we choose the initial conditions for $\dot{C}_2$ by assuming that the initial effective equation of state is close to $w_\textrm{eff}\approx-1.2$. After setting $C_2(t_0)$, the initial conditions for $Q$ and $\dot{Q}$ can be determined by Eqs.~\eqref{qeqconnect2} and \eqref{rhoconnec2}.

\begin{figure}[t]
\centering
\includegraphics[width=300pt]{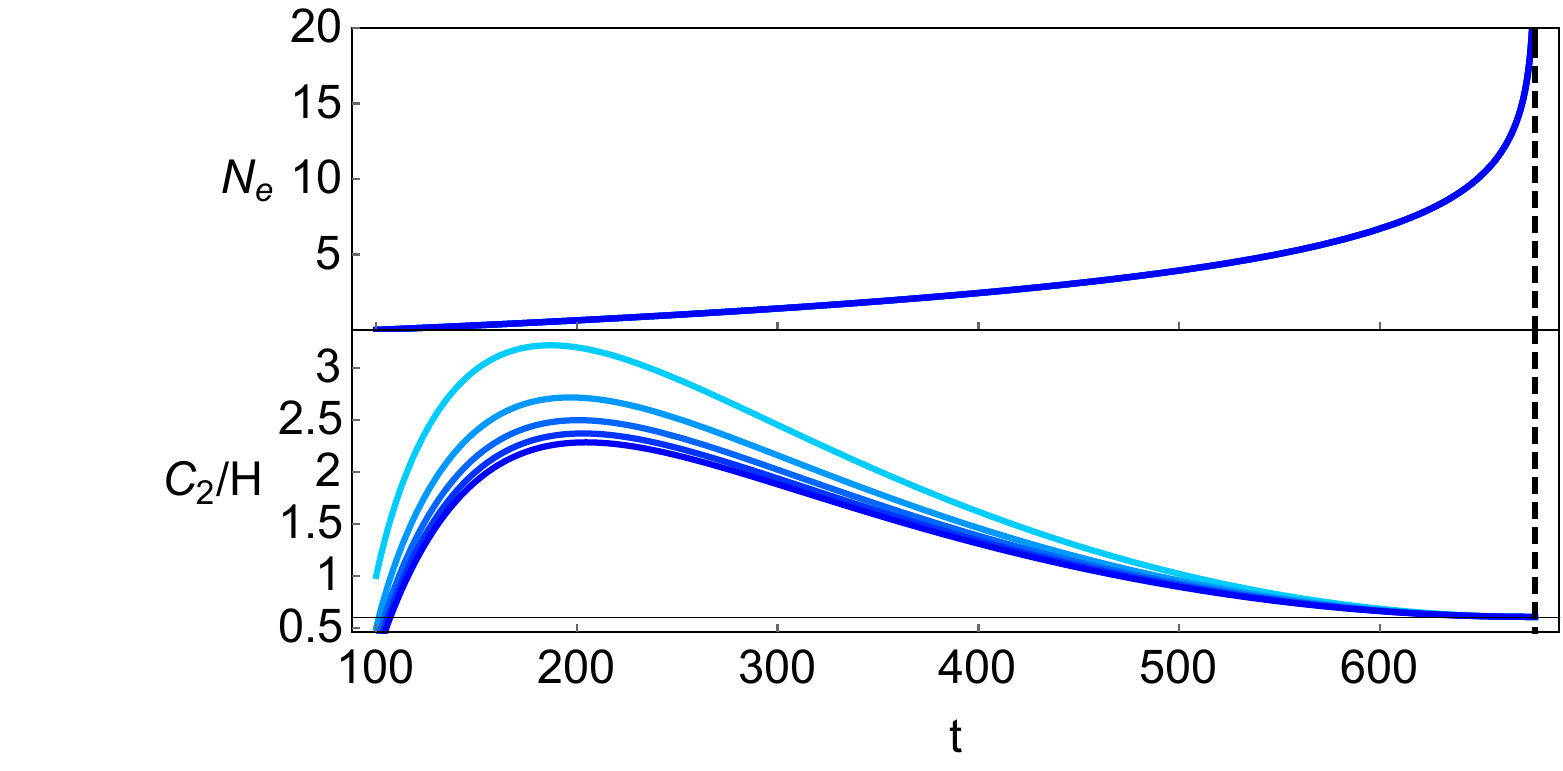}
\caption{The e-folds $N_e(t)$ (top) and the ratio $C_2/H$ (bottom) as functions of cosmic time $t$ for the phantom-dominated cosmology with connection $\Gamma ^{\text{(II)}}_Q$. The vertical dashed lines represent the time of the Big Rip singularity. The initial conditions  are chosen as $H(t_0)/C_2(t_0)=5$, $4$, $3$, $2$, and $1$, from blue to cyan.}
\label{fig:con2bigrip}
\end{figure}

In Fig.~\ref{fig:con2bigrip}, we exhibit the numerical results for the cosmological solutions with the initial conditions $H(t_0)/C_2(t_0)=5$, $4$, $3$, $2$, and $1$ (from blue to cyan). Similar to the case of $\Gamma ^{\text{(I)}}_Q$, the non-metricity $Q$ quickly converges to a non-zero constant when $t$ grows. Hence, the solutions approach GR-like and end up with a Big Rip singularity (the dashed vertical lines) where $w_\textrm{eff}\rightarrow w_0$. In addition, following the same procedure as what we did to obtain Eq.~\eqref{c3hfinal1}, we find that near the singularity, the ratio $C_2/H$ converges to
\be
\frac{C_2}{H}\rightarrow\frac{4}{3(1-w_0)}\,,
\ee
as indicated by the horizontal line in the bottom panel of Fig.~\ref{fig:con2bigrip}. Note that, unlike the case of $\Gamma ^{\text{(I)}}_Q$ where the Big Rip singularity can be delayed significantly (see Fig.~\ref{fig:con1bigrip}), the singularity in the case here can only be delayed very mildly by decreasing the value of $H(t_0)/C_2(t_0)$, which cannot be identified in Fig.~\ref{fig:con2bigrip}.

Let us summarize what we have demonstrated in sec.~\ref{sec:fqcosmology}. By considering the Born-Infeld $f(Q)$ gravity, i.e., Eq.~\eqref{BIf}, and assuming separately a radiation-dominated and a phantom-dominated universe, we show that the behaviors of cosmological solutions and, in particular, the regularity of the solutions, strongly depend on the choice of the connections. For the connection $\Gamma ^{\text{(III)}}_Q$, the bound on the non-metricity $Q$ implies directly that the Hubble function $H$ is bounded from above. Therefore, both the Big Bang and the Big Rip singularities are replaced by de Sitter phases. However, for connections $\Gamma ^{\text{(I)}}_Q$, the singularities remain. In particular, the solutions possess some interesting properties:
\begin{itemize}
\item The Big Bang singularity is governed by a stiff-like effective matter with the values of $w_\textrm{eff}$ at the singularity varying with the choices of initial conditions.
\item The universe may cross the phantom divide line before reaching the Big Rip singularity.
\end{itemize}
Finally, for connections $\Gamma ^{\text{(II)}}_Q$, both types of singularities also remain. In particular, the solutions converge to GR-like very quickly both toward the past and the future, hence the solutions remain almost GR-like. 

Before closing this section, we would like to mention that the existence of the Big Bang and Big Rip singularities in the cosmological solutions we found for the connections $\Gamma ^{\text{(I)}}_Q$ and $\Gamma ^{\text{(II)}}_Q$ are quite robust against different choices of initial conditions at $t_0$, as long as the cosmological behaviors at $t_0$ are close to that in standard GR cosmology. However, we still cannot completely exclude the possibility that there may be some fine-tuned initial conditions such that the singularities are resolved for these connections. A thorough analysis of the dependence between the choices of initial conditions and the resulting cosmological solutions can be carried out efficiently via dynamical analysis. We will leave this for future work. 

\section{Conclusions}\label{sec:conclustion}

In this work, we have conducted an extensive investigation into the cosmological implications of modified gravity within the framework of $f(Q)$ theories — gravitational models constructed in the context of symmetric teleparallel geometry, where the gravitational interaction is governed by non-metricity rather than curvature or torsion. A central theme of the study is the exploration of how the choice of affine connection, even under the constraints of symmetric teleparallel gravity (zero curvature and zero torsion), significantly influences the resulting cosmological dynamics.

We rigorously analyze three distinct types of connections — labeled as $\Gamma^{\text{(I)}}_Q$, $\Gamma^{\text{(II)}}_Q$, and $\Gamma^{\text{(III)}}_Q$ — which are compatible with the homogeneity and isotropy of the FLRW metric. Despite being geometrically equivalent under the flatness and torsionless conditions of symmetric teleparallel gravity, these connections lead to distinct physical predictions in cosmological scenarios, especially when extended to nontrivial forms of $f(Q)$.

Through both analytical and numerical methods, our results highlight the deep connection-dependence of cosmological behavior. In the analytical part, we derive conditions under which maximally symmetric (de Sitter) spacetimes arise, both in vacuum and in the presence of matter fields. The results show that while all three connections allow de Sitter solutions in vacuum, their behavior diverges sharply when matter is introduced. Notably, the third connection $\Gamma^{\text{(III)}}_Q$, severely restricts the allowed matter content, supporting only vacuum or cosmological constant configurations, for regular $f''(Q)$.

Further analytical explorations into specific functional forms of $f(Q)$, such as power-law and exponential models, demonstrate the possibility of reconstructing cosmological solutions by tuning the free function $f(Q)$ or the connection parameters. Particularly, the form of $f(Q)$ can be tailored to yield scale factors mimicking early-universe inflationary expansion or late-time accelerated expansion, without invoking exotic matter fields.

Our numerical section delves into the Born-Infeld form of $f(Q)$, motivated by its potential to regularize singularities due to the bounded nature of the non-metricity scalar $Q$. Here, the importance of the connection choice is especially prominent. For example, with connection $\Gamma^{\text{(III)}}_Q$, a radiation-dominated universe avoids the traditional Big Bang singularity, transitioning instead into a non-singular early de Sitter phase. Likewise, a phantom-dominated universe under the same connection avoids the Big Rip, asymptotically approaching a de Sitter state. However, for the other connections $\Gamma^{\text{(I)}}_Q$ and $\Gamma^{\text{(II)}}_Q$, the singularities persist, albeit with nuanced behaviors. In the case of $\Gamma^{\text{(I)}}_Q$, the early universe shows stiff-matter-like behavior near the singularity, and the late-time evolution can involve oscillations in the effective equation of state, including phantom-divide crossings. The second connection $\Gamma^{\text{(II)}}_Q$, while similar to general relativity in many aspects, shows limited capability to alter singular behavior, with its cosmological evolution quickly converging to GR-like solutions. It is interesting to notice that although at the Big Bang and Big Rip singularities the metric is not well defined for any of the models that we have analysed in the previous section, the geometric term $Q$ remains finite despite involving the derivative of the metric. This is probably due to the linear combination used to define the scalar $Q$ in Eqs.~\eqref{Q91} and \eqref{qeqconnect2}. As long as $C_2$ and $C_3$ are dynamical, the finiteness of $Q$ does not imply a regular behavior of the metric.

Our study underscores a key insight: the affine connection in $f(Q)$ theories is not merely a mathematical artifact but a physically relevant structure that can lead to distinct and measurable cosmological consequences, especially near singular regimes. Therefore, future efforts in modified gravity must not overlook the role of the connection, as it opens pathways to novel physics without invoking additional degrees of freedom or exotic matter. This paper lays a solid foundation for deeper investigations into the role of geometry in shaping the universe’s past and future, and highlights the untapped potential of $f(Q)$ gravity in addressing unresolved cosmological puzzles such as singularity avoidance, inflation, and dark energy phenomenology.

\section*{Acknowledgments}

We thank José Beltrán Jiménez and Erik Jensko for their enlightening discussions and valuable feedback. This article is based upon work from COST Action CA21136 Addressing observational tensions in cosmology with systematics and fundamental physics (CosmoVerse) supported by COST (European Cooperation in Science and Technology). IA is supported by the Basque government Grant No. IT1628-22 (Spain), by Grant PID2021-123226NB-I00 (funded by MCIN/AEI/10.13039/501100011033 and by “ERDF A way of making Europe”). MBL is supported by the Basque Foundation of Science Ikerbasque. Her work is also supported by the Spanish grant PID2023-149016NB-I00 (funded by MCIN/AEI/10.13039/501100011033 and by “ERDF A way of making Europe") and the Basque government Grant No. IT1628-22 (Spain). MBL acknowledges as well the hospitality of Yangzhou University and Jiangsu University of Science and Technology, where this work was initiated. CYC is supported by the Special Postdoctoral Researcher (SPDR) Program at RIKEN. CYC also thanks APCTP, Pohang, Korea, for their hospitality during the External Program APCTP-GW2025, from which this work greatly benefited. KFD's work was supported by the PNRR-III-C9-2022–I9 call, with project number 760016/27.01.2023. XYC is supported by Jiangsu University of Science and Technology (JUST).

\bibliography{bib}

\end{document}